\documentclass[11pt,a4paper]{article}

\usepackage{jheppub}
\usepackage{psfrag}
\usepackage{slashed}
\usepackage{xspace}
\usepackage{subfigure}
\usepackage{pslatex}

\usepackage{hyperref}
\hypersetup{bookmarksnumbered,}

\usepackage{bbm}
\usepackage{framed}
\usepackage{verbatim}
\usepackage{multirow}

\graphicspath{{figures/}}

\newcommand{\ngluon}{{\sc NGluon}\xspace}
\newcommand{\NJet}{{\sc NJet}\xspace}
\newcommand{\BlackHat}{{\sc BlackHat}\xspace}

\definecolor{mygreen}{rgb}{0,0.7,0}

\newcommand{\eps}{\epsilon}

\newcommand{\qbar}{{\overline{q}}}
\newcommand{\dbar}{{\overline{d}}}
\newcommand{\ubar}{{\overline{u}}}
\newcommand{\sbar}{{\overline{s}}}

\newcommand{\AmpF}[6]{A^{[f]}\!(#1#2#3#4#5#6)}

\def\Ref#1{Ref.~\cite{#1}}
\preprint{HU-EP-12/29}

\title{Numerical evaluation of virtual corrections to multi-jet production in massless QCD}

\author[a]{Simon Badger}
\author[b]{Benedikt Biedermann}
\author[b]{Peter Uwer}
\author[a]{Valery Yundin}
\affiliation[a]{
Niels Bohr International Academy and Discovery Center, The Niels Bohr Institute,\\%
University of Copenhagen, Blegdamsvej 17, DK-2100 Copenhagen, Denmark}
\affiliation[b]{
Humboldt-Universit\"at zu Berlin, Institut f\"ur Physik,\\%
Newtonstra{\ss}e 15, D-12489 Berlin, Germany}

\emailAdd{badger@nbi.dk}
\emailAdd{benedikt.biedermann@physik.hu-berlin.de}
\emailAdd{peter.uwer@physik.hu-berlin.de}
\emailAdd{yundin@nbi.dk}

\abstract{ We present a C++ library for the numerical evaluation of
  one-loop virtual corrections to multi-jet production in massless
  QCD. The pure gluon primitive amplitudes are evaluated using \ngluon
  \cite{Badger:2010nx}. A generalized unitarity reduction algorithm is
  used to construct arbitrary multiplicity fermion-gluon primitive
  amplitudes.  From these basic building blocks the one-loop
  contribution to the squared matrix element, summed over colour and
  helicities, is calculated. No approximation in colour is performed.
  While the primitive amplitudes are given for arbitrary
  multiplicities we provide the squared matrix elements only for up to
  7 external partons allowing the evaluation of the five jet cross
  section at next-to-leading order accuracy. The library has been
  recently successfully applied to four jet production at next-to-leading order in
  QCD \cite{Badger:2012zz}.
  \vspace{\baselineskip}\\
  Draft from \today }


\makeatletter\gdef\@fpheader{~}\makeatother
\begin{document}

\maketitle
\flushbottom

\section{Introduction}

Next-to-leading order (NLO) corrections to multi-particle Standard Model processes at hadron
colliders are essential for quantitative predictions of the large QCD background to many new physics
scenarios.

Though the computation of the one-loop virtual corrections with four or more final state particles
has always presented a theoretical challenge, new methods like unitarity
\cite{Bern:1994zx,Bern:1994cg}, generalized unitarity
\cite{Britto:2004nc,Ellis:2007br,Giele:2008ve,Forde:2007mi} and integrand reduction
\cite{Ossola:2006us} have been developed into effective tools suited for a wide variety of
processes. We also note that the traditional approach---the powerful work
horse of the last three decades for perturbative calculations in
quantum field theory---has been refined and improved
continuously. For recent developments we refer for example to
\Ref{Cascioli:2011va}. 

A number of automated approaches have been developed in the last five years
\cite{Ossola:2007ax,Giele:2008bc,Berger:2008sj,Mastrolia:2010nb,Hirschi:2011pa,Cullen:2011ac} some
of which are now publicly available. On-shell techniques have also been used recently to compute compact
analytic matrix elements within the popular MCFM program
\cite{Campbell:2012dh,Badger:2010mg,Badger:2011yu,Campbell:2012uf,Campbell:2010cz,Campbell:2011cu}.
The availability of one-loop corrections combined with real radiation contributions have made an increasingly
large number of phenomenological studies possible
\cite{Berger:2009zg,Berger:2009ep,Berger:2010vm,Bredenstein:2009aj,Bredenstein:2010rs,Bevilacqua:2009zn,
Campanario:2011ud,Bevilacqua:2010ve,Denner:2010jp,Bevilacqua:2010qb,Bevilacqua:2011aa,Denner:2012yc,Bevilacqua:2012em,
Binoth:2009rv,Greiner:2011mp,Greiner:2012im,Melnikov:2009wh,KeithEllis:2009bu,Melia:2011dw,Melia:2010bm,
Berger:2010zx,Ita:2011wn,Frederix:2010ne,Becker:2011vg}. More recently
improvements beyond fixed-order in perturbation theory using parton shower matching techniques
\cite{Frixione:2002ik,Frixione:2007vw,Alioli:2010xd} have been
achieved for a number of non-trivial cases \cite{Alioli:2010xa,Garzelli:2011is,Garzelli:2011vp,Frederix:2011zi,Frederix:2011qg,Kardos:2011na,Kardos:2011qa,Alioli:2011as,Campbell:2012am,Hoeche:2012ft}.
Despite the tremendous progress there is still room for improvement in terms of
efficiency and flexibility before one-loop computations reach the same level of automation as tree-level predictions
and a number of interesting new directions continue to appear \cite{Giele:2009ui,Cascioli:2011va,Becker:2012aq}.

In this paper we present the C++ library \NJet for the numerical evaluation of one-loop amplitudes for multi-jet
production in massless QCD. The on-shell generalized unitarity
algorithm employed in \NJet is well suited to
deal with the large amount of gluonic radiation which complicates a traditional Feynman approach to
such processes. Our paper is organized as follows: first we briefly review the method of
generalized unitarity used for the construction of primitive amplitudes with arbitrary numbers of
fermion pairs. We then describe the construction of full colour and helicity summed squared
amplitudes by partial amplitudes in terms of the computed primitive
amplitudes. A detailed description of the installation procedure and
the usage of the \NJet package is presented in section
\ref{sec:install} including a short summary of the interface
via the Binoth Les Houches Accord (BLHA)~\cite{Binoth:2010xt}. In
section \ref{sec:results} we discuss the validation of the \NJet
library. In particular we present cross checks with available
analytic results and show a numerical comparison  for selected
benchmark points. In addition we present some performance tests of
the program that investigate numerical accuracy and speed of both primitive and colour summed amplitudes.

\section{Methods}

The \NJet library is built upon the framework of \ngluon \cite{Badger:2010nx} which uses generalized
unitarity to construct primitive gluon amplitudes. This  procedure allows one-loop amplitudes
to be constructed from tree-level amplitudes via the solution to a system of cut equations.
One important aspect of this method is that it maps the problem of computing any one-loop amplitude
to the solution of a linear system of equations which is well suited
for numerical evaluation. Furthermore the method uses on-shell
amplitudes instead of individual Feynman diagrams avoiding the
significant growth in complexity when one-loop amplitudes for large
multiplicities are considered.

The implementation in \ngluon uses the algorithm of Ellis, Giele, Kunszt and Melnikov
\cite{Ellis:2007br,Giele:2008ve} to construct the coefficients of the
scalar one-loop integrals. For the
rational terms we map the problem of cuts in $4-2\eps$ dimensions to that of four dimensional
massive cuts. The scalar one-loop integrals are all known analytically and
are evaluated using the \texttt{FF/QCDLoop} package \cite{vanOldenborgh:1990yc,Ellis:2007qk}. An alternative
implementation by van Hameren now allows for full complex masses and quadruple precision evaluation
of the integral functions \cite{vanHameren:2010cp}.

\subsection{Multiple Fermion Primitive Amplitudes}\label{sec:multiFermPrims}

\begin{figure}[b!]
  \begin{center}
    \includegraphics[width=\textwidth]{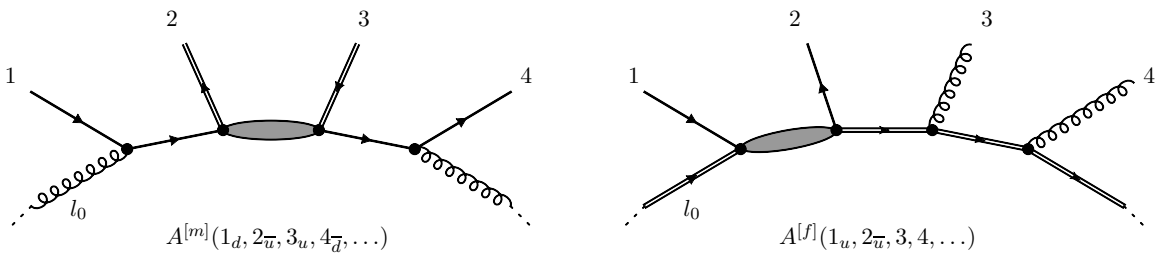}
  \end{center}
  \caption{Constructing parent diagrams for the primitive
    amplitudes. In case that no internal propagator can be connected
  since the corresponding vertices do not exist a blank propagator is
  inserted, represented graphically as a shaded blob.}
  \label{fig:prim}
\end{figure}

A new feature added to the \ngluon package to enable the application to multi-jet production at NLO
is the ability to compute primitive amplitudes with an arbitrary number of external and internal
quarks. First results have been presented recently in references
\cite{Badger:2011zv,Badger:2012dd}. In the following we briefly
comment on this extension of \ngluon.

An $n$-particle, one-loop primitive amplitude is defined by an ordered set of $n$ external
particles and a set of propagators specifying the loop content.
These primitive amplitudes are gauge invariant sums of colour ordered Feynman diagrams as introduced
in Ref.~\cite{Bern:1994fz}. The full set of independent primitives falls into two subclasses:
\begin{itemize}
  \item Amplitudes with a mixture of quark and gluon propagators, $A^{[m]}_n$,
  \item Amplitudes with an internal fermion loop, $A^{[f]}_n$.
\end{itemize}
Each primitive amplitude can be associated uniquely with a ``parent
diagram" \cite{Ellis:2011cr}.  This is a diagram with $n$ propagators
and $n$ colour ordered three-vertices from which all contributing
ordered Feynman diagrams can be obtained by ``pinching" and ``pulling"
operations. The parent diagrams of the two classes of primitives above
are determined by the external legs and the parton type of the
propagator appearing immediately before the first leg. 
All other internal parton types will then follow from the preceding
ones, two examples of this are shown in Figure~\ref{fig:prim}.  In
cases where the corresponding vertices do not exist and no internal
propagator can be connected we attribute a special propagator as
placeholder in order to ensure that the parent diagram always contains
a total of $n$ propagators.

Once the parent diagram of the primitive amplitude has been constructed the generalized unitarity
algorithm can be applied to compute the coefficients of the scalar integral basis by factorizing
each cut into a product of tree-level amplitudes. These amplitudes are then computed with
off-shell Berends-Giele recursion relations \cite{Berends:1987me}.  With a suitable caching system for the
off-shell currents, this approach scales polynomially in complexity at $O(n^4)$ with $n$ being the
number of external legs.  The power of four arises from the four-gluon vertex, which dominates the
recursion for large $n$, independent on the number of quark pairs.  The inclusion of fermions is
rather a matter of book-keeping and no conceptual difficulties
arise. 

The rational terms are evaluated by using four-dimensional massive cuts to obtain $D$-dimensional
cut information \cite{Bern:1995db,Bern:1996ja}. This corresponds to working in the four-dimensional
helicity (FDH) scheme \cite{Bern:1991aq} which is equivalent to dimensional reduction at one-loop.
The coefficients of the $D$-dimensional integral basis \cite{Giele:2008ve} are efficiently extracted
using a discrete Fourier projection on the internal mass \cite{Badger:2008cm,Berger:2009zb}. When
only concerned with the rational terms it is possible to replace all internal $D$-dimensional gluons
with massive scalars and treat $D$-dimensional fermions with massive fermions. The computation of
the rational term is therefore a direct copy of the extraction of the integral coefficients but with
different tree-level amplitudes which are also straightforward to compute with off-shell recursion
relations.  In addition one can use supersymmetry identities \cite{Bern:1994fz} to find relations between the rational
terms of the two classes of primitives $A^{[m]}$ and $A^{[f]}$ which allows for a more efficient
construction of the full colour amplitudes. This has been implemented for fermion loops in both the
pure gluon and single fermion pair amplitudes:
\begin{align}
  &\mathcal{R}\left( A_n^{[m]}(1,\ldots,n) - A_n^{[f]}(1,\ldots,n) \right) = 0, \\
  &\mathcal{R}\Big(
  A_n^{[m]}(1_\qbar,2,\ldots,(k-1),k_q,(k+1)\ldots,n) \nonumber\\&
  + (-1)^n A_n^{[m]}(1_\qbar,n,\ldots,(k+1),k_q,(k-1),\ldots,2) \nonumber\\&
  - A_n^{[f]}(1_\qbar,2,\ldots,(k-1),k_q,(k+1)\ldots,n) \nonumber\\&
  -(-1)^n A_n^{[f]}(1_\qbar,n,\ldots,(k+1),k_q,(k-1),\ldots,2)
  \Big) = 0,
  \label{eq:SUSY}
\end{align}
where the operator $\mathcal{R}$ denotes taking the rational part of the amplitude.

\subsection{Partial Amplitudes and Colour Summation}

To build the full colour amplitudes from the primitive one-loop amplitudes provided by \ngluon and
its extension to massless quarks we must construct the coefficients of the $SU(N_c)$ colour factors,
the partial amplitudes.  In general the full colour $L$-loop amplitude can be written as:
\begin{equation}
  \mathcal{A}_n^{(L)}(\{p_i\})  = \sum_{c} T^{(L)}_c\!(\{a_i\})\, A_{n;c}^{(L)}(p_1,\ldots,p_n),
  \label{eq:gencolouramp}
\end{equation}
where $A_{n;c}$ are the partial amplitudes and the summation is performed over
all different colour structures $T^{(L)}_c\!(\{a_i\})$
which depend on the colour of the external partons abbreviated here
by~$a_i$. 
For the evaluation of the colour summed squared matrix elements we
construct colour matrices $\mathcal{C}$ by taking the product of colour structures and summing over the colour degrees of
freedom of the external partons,
\begin{align}
  \mathcal{C}^{(0)}_{cc'} &= \sum_{i} T^{(0)}_{c}\!(\{a_i\}) \cdot T^{(0)}_{c'}\!(\{a_i\}),\\
  \mathcal{C}^{(1)}_{cc'} &= \sum_{i} T^{(0)}_{c}\!(\{a_i\}) \cdot T^{(1)}_{c'}\!(\{a_i\}).
\end{align}
The colour summed squared Born amplitude will then be constructed from the tree-level partial amplitudes
$A_{n;c}^{(0)}$ according to
\begin{align}
  d\sigma^{LO}  \sim A_{n;c}^{(0)\dagger} \cdot \mathcal{C}^{(0)}_{cc'} \cdot A_{n;c'}^{(0)}.
  \label{eq:colsumLO}
\end{align}
The one-loop interference term is given in complete analogy by
\begin{align}
  d\sigma^{NLO}  \sim A_{n;c}^{(0)\dagger} \cdot \mathcal{C}^{(1)}_{cc'} \cdot A_{n;c'}^{(1)}.
  \label{eq:colsumNLO}
\end{align}
The entries of the colour matrix are given in terms of products of
$SU(N_c)$ generators summed over colour indices which we evaluate
directly into a rational function in $N_c$ before hard coding the
result into~C++.

\begin{figure}[t]
  \begin{center}
    \includegraphics[width=\textwidth]{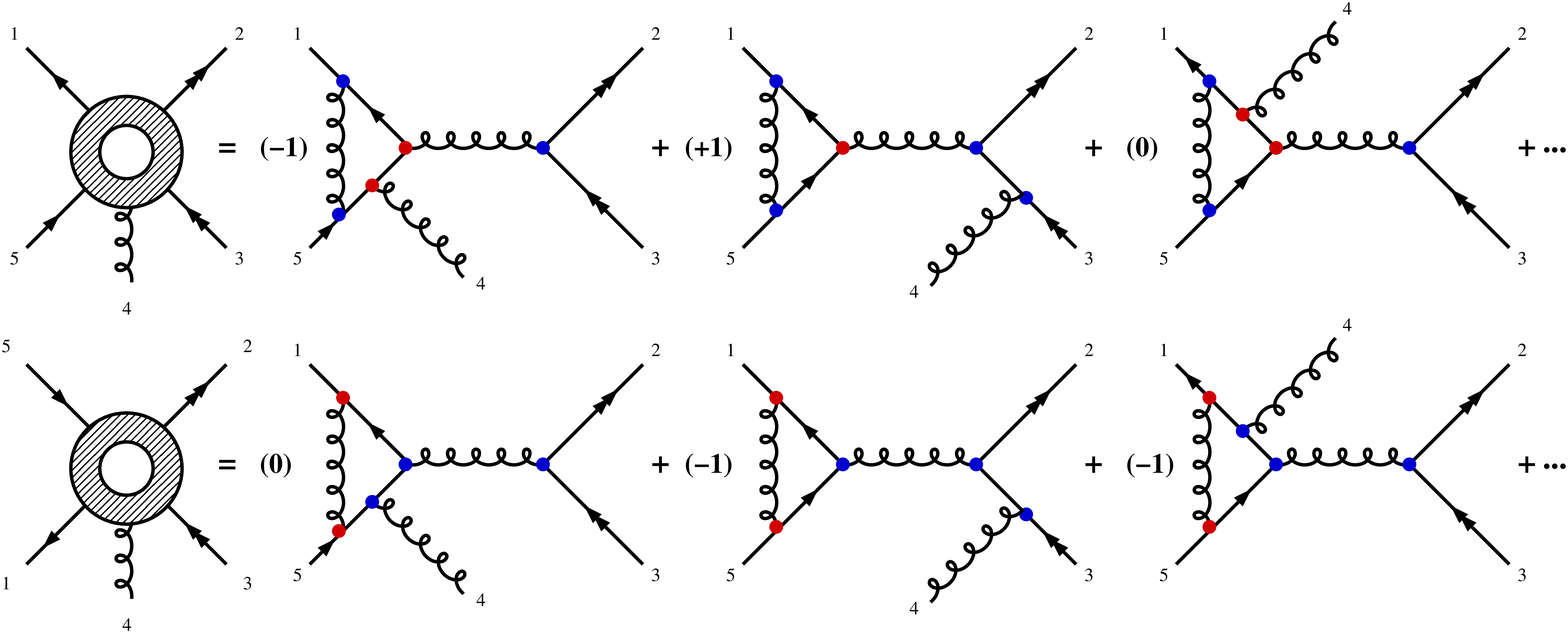}
  \end{center}
  \caption{Matching Feynman diagrams to the primitive amplitudes
  $A_5^{[m]}(1_\dbar,2_u,3_\ubar,4_g,5_d)$ and $A_5^{[m]}(1_\dbar,4_g,3_\ubar,2_u,5_d)$.
  Each vertex is either ordered (red) or unordered (blue) with respect to the primitive in question.
  A diagram contributes with a coefficient $-1$ when it contains an odd number of unordered vertices
  and $+1$ otherwise. Any diagram which cannot be obtained from the original primitive
  by ``pinching" and ``pulling" operations will have a $0$ coefficient.
  }
  \label{fig:matching}
\end{figure}

The algorithm for determining the decomposition of the partial
amplitudes in terms of primitive amplitudes has been constructed along
the lines of that suggested by Ellis, Giele, Kunszt, Melnikov and
Zanderighi \cite{Ellis:2008qc,Ellis:2011cr} and that of Ita and Ozeren
\cite{Ita:2011ar}. The principle is rather simple:

\begin{enumerate}
\item First we generate all the Feynman diagrams $D_i$ for a given
  $n$-parton amplitude, separating each one into a colour factor
  denoted by $C_i$ and a kinematic factor containing the Lorentz
  structure denoted by $K_i$:
    \begin{equation}\label{eq:colorAmp}
      \mathcal{A}_n = \sum_{i=1}^{N_\text{dia}} D_i =
      \sum_{i=1}^{\hat{N}_\text{dia}} C_i K_i, \qquad 
    \end{equation}
    where $N_\text{dia}$ is the total number of diagrams and
    $\hat{N}_\text{dia}$ is the number of diagrams excluding ones with
    four point interactions\footnote{The kinematic parts $K_i$ are
    understood to contain 
      contributions from both $3$-gluon and $4$-gluon vertices.}. Each
    diagram's colour factor $C_i$ is a known linear combination of
    trace factors $T_c(\{a\})$ with coefficients $F_{ci}$:
    \begin{equation}
      C_i = \sum_c
      T_c(\{a\})\, F_{c i}.
    \end{equation}
    For clarity
    we omit the loop superscript on the colour dressed amplitude
    $\mathcal{A}_n$ for the remainder of this section.
  \item Secondly we write all possible primitive amplitudes $P_i$ as combinations
    of colour-stripped diagrams:
    \begin{gather}\label{eq:colorM}
      P_i = \sum_{j=1}^{\hat{N}_\text{dia}} M_{ij} K_j, \qquad i \in \{1,2,\ldots,N_\text{pri}\}, \qquad
      N_\text{pri} = N^{[m]}_\text{pri} + N^{[f]}_\text{pri}\\
      N^{[f]}_\text{pri} = (n-1)!, \qquad
      N^{[m]}_\text{pri} = \begin{cases}
        (n-1)!, & \text{$n_q=0$} \\
        \frac{1}{2}n_q(n-1)!, & \text{$n_q=2,4,\ldots$}
      \end{cases}
    \end{gather}
    where $n_q$ is the number of external quarks. Obviously the set
    $\{P_i\}$ is overcomplete and $N_\text{pri}$ is only an upper
    bound for the number of primitives.  The coefficients $M_{ij}$
    have entries of $\pm 1$ and $0$ depending on how the parent
    diagram representing the ordered primitive amplitude $P_i$ matches
    onto the kinematic topology of $K_j$ ($1$ -- even-ordered, $-1$ --
    odd-ordered, $0$ -- no match).  Figure \ref{fig:matching} depicts
    an example of the matching for the primitive amplitude
    $A_5^{[m]}(1_\dbar,2_u,3_\ubar,4_g,5_d)$, the notion of odd and
    even is taken with respect to the colour ordered Feynman rules
    (see for instance \Ref{Dixon:1996wi}).

    If some diagrams are known to have additional non-trivial relations, this information
    can be included at this step to reduce the number of columns in $M_{ij}$.
    For instance Furry's theorem can be used on fermionic loops to obtain additional simplifications.

  \item The rank of the matrix obtained at the previous step defines the number of
        independent primitive amplitudes, which is usually much smaller than the initial upper bound.
    \begin{gather}
      \hat N_\text{pri} = \operatorname{rank}M,
      \quad \hat N_\text{pri} \leq N_\text{pri},
      \quad \hat N_\text{pri} \leq \hat{N}_\text{dia}
    \end{gather}

    To solve the equation \eqref{eq:colorM}
    it is convenient to consider the augmented matrix $\hat M = [M | {-}\mathbbm{1}]$,
    where $\mathbbm{1}$ is the $N_\text{pri}$-dimensional identity matrix.
    By putting it into reduced row echelon form (RREF), we get a solution of \eqref{eq:colorM}
    as the upper $\hat N_\text{pri}$ rows of $\operatorname{RREF}( \hat M)$.
    The lower $N_\text{pri}-\hat N_\text{pri}$ rows will contain the left null space of the matrix~$M$,
    which encodes relations between primitive amplitudes.
    A solution with a minimal basis of primitives is guaranteed by the properties of the RREF operation.
    \begin{equation}\label{eq:colorSol}
      K_i = \sum_{j=1}^{\hat N_\text{pri}} B_{ij} \hat P_j,
      \qquad \{\hat P_j\}_{\hat N_\text{pri}} \subset \{ P_j\}_{\vphantom{\hat N}N_\text{pri}}.
    \end{equation}
  \item After substituting solution \eqref{eq:colorSol} into \eqref{eq:colorAmp} and expressing the
    colour factors $C_i$ using the trace basis $T_c$ we can form the partial colour
    amplitudes in terms of independent primitive amplitudes $\hat P_i$ as in \eqref{eq:gencolouramp},
    \begin{equation}\label{eq:colorAns}
      \mathcal{A}_n =
      \sum_c T_c(\{a\})\, A_{n;c}
      = \sum_c T_c(\{a\}) \sum_{j=1}^{\hat N_\text{pri}} Q_{c j} \hat P_j,
      \qquad Q_{c j} = \sum_{i=1}^{\hat{N}_\text{dia}} F_{ci} B_{ij}.
    \end{equation}
\end{enumerate}
We are now free to use the result of the primitive decomposition, eq. \eqref{eq:colorAns}, to
evaluate the colour sums in eqs. \eqref{eq:colsumLO} and \eqref{eq:colsumNLO} in terms of primitive
amplitudes.

Though this procedure involves generation of Feynman diagrams, only 
topologies of graphs are required making the application even to a relatively high multiplicity straightforward.
Nevertheless the generic algorithm outlined above can be further improved.
The four gluon vertex carries no additional information as far as colour is concerned and
can be ignored leading to a substantial speed up.
In general, since we are only looking for one specific solution of \eqref{eq:colorM},
any subset of diagrams which results in the matrix with the same rank as $M_{ij}$
will be sufficient to obtain the correct form of \eqref{eq:colorAns}.

The number of independent primitive amplitudes for the pure gluon and single fermion pair channels
follows the well known structure outlined in references \cite{Bern:1994zx,Bern:1994fz}.
Our approach to the fermion loops exploiting Furry type symmetries differs from that of Ita and Ozeren
\cite{Ita:2011ar} explaining the smaller number of primitive amplitudes we obtain in the
multi-fermion channels. Table~\ref{tab:primitivebasis} lists the number of primitives for all
channels required for $\leq 5$ jet production.
As an example we show the fermion loop contribution to the six-quark
amplitude
\def\mixed{\mbox{\scriptsize mixed}}
\def\qloop{q\mbox{\scriptsize -loop}}
\begin{displaymath}
 A_{6q}(1_{\bar u}, 2_{\vphantom{\bar u}u}, 3_{\bar d}, 4_{\vphantom{\bar d}d}, 5_{\bar s}, 6_{\vphantom{\bar s}s}) = A_{6q}^{\mixed} + n_f A_{6q}^{\qloop} 
\end{displaymath}
expressed in terms of 4 primitive amplitudes:
\begin{multline*}
A_{6q}^{\qloop} =   \frac{1}{N_c} \Bigl(T_3 + T_4 - T_1 N_c - \frac{1}{N_c} T_6 \Bigr)
\times \AmpF123456
+ \frac{1}{N_c}\left(T_5 - T_3\right) \times\AmpF124365 \\
+ \frac{1}{N_c}\Bigl(2 T_5 - T_2 N_c - \frac{1}{N_c}T_6 \Bigr)\times \AmpF125634 
+ \frac{1}{N_c}\left(T_5 - T_4\right) \times \AmpF126534
\end{multline*}
where $T_i$ are the basis of colour factors
\begin{gather*}
\begin{aligned}
T_1 &= \delta_{1 6}\delta_{3 2}\delta_{5 4}, \qquad &
T_2 &= \delta_{1 4}\delta_{3 6}\delta_{5 2}, \qquad &
T_3 &= \delta_{1 2}\delta_{3 6}\delta_{5 4}, \\
T_4 &= \delta_{1 4}\delta_{3 2}\delta_{5 6}, &
T_5 &= \delta_{1 6}\delta_{3 4}\delta_{5 2}, &
T_6 &= \delta_{1 2}\delta_{3 4}\delta_{5 6}.
\end{aligned}
\end{gather*}

\begin{table}[h]
\centering
  \begin{tabular}[h]{|l|c|c|c|}
  \hline
  & $A^{[0]}$ & $A^{[m]}$ & $A^{[f]}$ \\
  \hline
  $\mathcal{A}_4(g,g,g,g)$ & 2 & 3 & 3 \\
  \hline
  $\mathcal{A}_4(\dbar,d,g,g)$ & 2 & 6 & 1 \\
  \hline
  $\mathcal{A}_4(\dbar,d,\ubar,u)$ & 1 & 4 & 1 \\
  \hline
\end{tabular}
\begin{tabular}[h]{|l|c|c|c|}
  \hline
  & $A^{[0]}$ & $A^{[m]}$ & $A^{[f]}$ \\
  \hline
  $\mathcal{A}_5(g,g,g,g,g)$ & 6 & 12 & 12 \\
  \hline
  $\mathcal{A}_5(\dbar,d,g,g,g)$ & 6 & 24 & 6 \\
  \hline
  $\mathcal{A}_5(\dbar,d,\ubar,u,g)$ & 3 & 16 & 3 \\
  \hline
\end{tabular}\\
\begin{tabular}[h]{|l|c|c|c|}
  \hline
  & $A^{[0]}$ & $A^{[m]}$ & $A^{[f]}$ \\
  \hline
  $\mathcal{A}_6(g,g,g,g,g,g)$ & 24 & 60 & 60 \\
  \hline
  $\mathcal{A}_6(\dbar,d,g,g,g,g)$ & 24 & 120 & 33 \\
  \hline
  $\mathcal{A}_6(\dbar,d,\ubar,u,g,g)$ & 12 & 80 & 13 \\
  \hline
  $\mathcal{A}_6(\dbar,d,\ubar,u,\sbar,s)$ & 4 & 32 & 4 \\
  \hline
\end{tabular}
\begin{tabular}[h]{|l|c|c|c|}
  \hline
  & $A^{[0]}$ & $A^{[m]}$ & $A^{[f]}$ \\
  \hline
  $\mathcal{A}_7(g,g,g,g,g,g,g)$ & 120 & 360 & 360 \\
  \hline
  $\mathcal{A}_7(\dbar,d,g,g,g,g,g)$ & 120 & 720 & 230 \\
  \hline
  $\mathcal{A}_7(\dbar,d,\ubar,u,g,g,g)$ & 60 & 480 & 75 \\
  \hline
  $\mathcal{A}_7(\dbar,d,\ubar,u,\sbar,s,g)$ & 20 & 192 & 20 \\
  \hline
\end{tabular}

\caption{The number of contributing independent primitive amplitudes at tree-level ($A^{[0]}$) and at one-loop for
the mixed ($A^{[m]}$) and fermion loop ($A^{[f]}$) cases. Like flavour amplitudes for multiple fermions
are obtained by symmetrization and therefore contain larger bases of primitives.}
\label{tab:primitivebasis}
\end{table}

In cases where the final state has a Bose symmetry (i.e. for gluons in the final state) the colour
summation can be further simplified. Since the phase space integration
is symmetric under permutations of the outgoing gluons, it is possible
to work with a `de-symmetrized' version of the squared matrix elements
and achieve the symmetrization through the phase space integration \cite{Ellis:2009zw},
\def\dLIPS{d\mbox{LIPS}}
\begin{align}
  \sigma^V_{gg\to n(g)} &
  = \int \dLIPS_n \,\,\sum_{c=1}^{(n-2)!} \sum_{c'=1}^{(n-1)!/2} A^{(0)\dagger}_{n;c} \cdot \mathcal{C}^{(1)}_{cc'} \cdot A^{(1)}_{n;c'} \nonumber\\&
  = \frac{(n-2)!}{2} \int \dLIPS_n \,\,\sum_{c=1}^{(n-2)!} \sum_{c'=1}^{n-1}  A^{(0)\dagger}_{n;c} \cdot \mathcal{C}^{(1), {\rm dsym}}_{cc'} \cdot A^{(1), {\rm dsym}}_{n;c'}
  \label{eq:dsymXS}
\end{align}
where $\dLIPS_n$ denotes the Lorentz invariant phase space measure of the $n$ parton final state.
In the case of the gluonic channels the saving is significant and brings the number of independent primitive amplitudes
from $\frac{1}{2}(n-1)!$ down to $(n-1)$. The de-symmetrized colour matrices $\mathcal{C}^{(1), {\rm dsym}}$ have
been computed and implemented in \NJet for all gluonic channels, the fermionic channels
however do not improve to the same extent and are implemented in the standard fashion.

We note that the number of independent primitives is significantly reduced in the case of the
leading colour approximation where additional symmetries can be exploited. A complete investigation
of the validity of such an approximation remains for a future study.

\subsection{Cache system for tree amplitudes \label{sec:cache}}

The various primitives contributing to a full-colour helicity summed squared matrix element differ
with respect to the external helicities, the permutations of the external legs and the loop content.
On the other hand, every primitive is constructed by sewing together partial tree-level amplitudes.
Therefore partial trees that agree in momenta, flavour and helicities of both external legs and
internal legs can be reused for different primitive amplitudes.

\begin{figure}[b]
  \centering
  \begin{minipage}[b]{0.45\textwidth}
    \includegraphics[width=\textwidth]{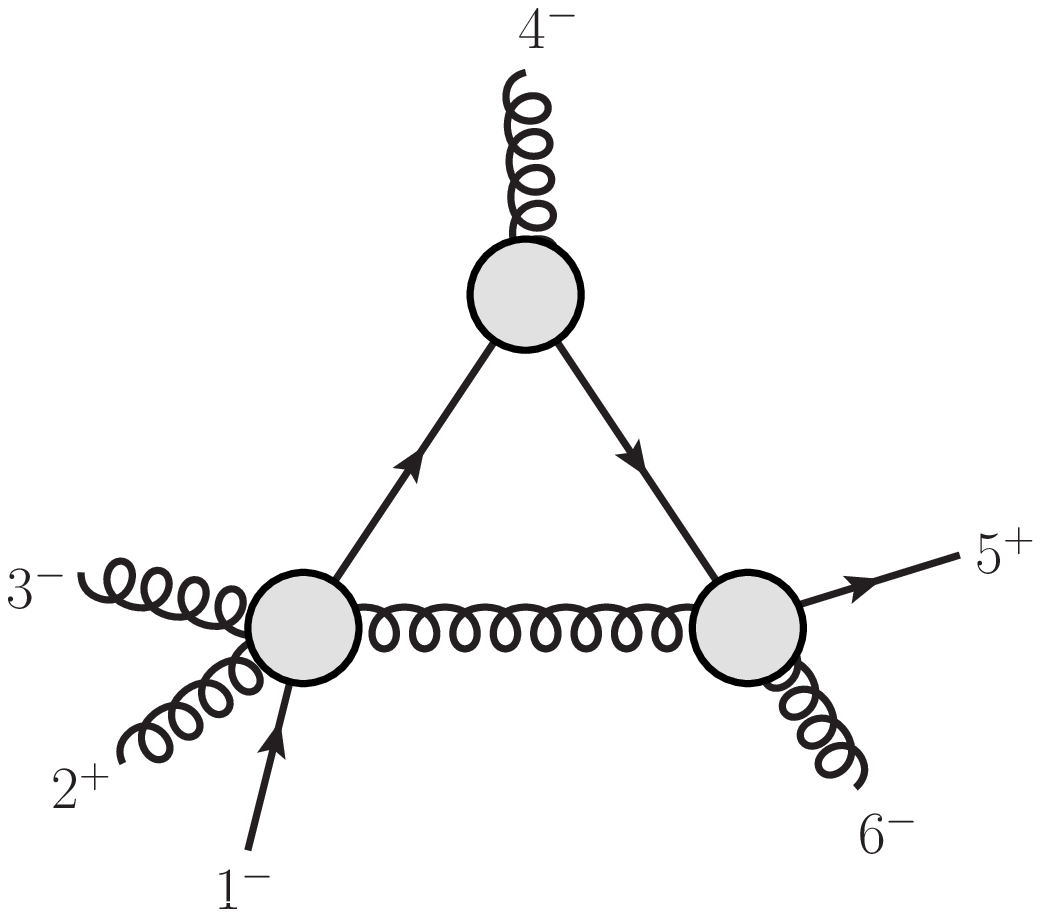}
  \end{minipage}
  \begin{minipage}[b]{0.45\textwidth}
    \centering
    \includegraphics[width=\textwidth]{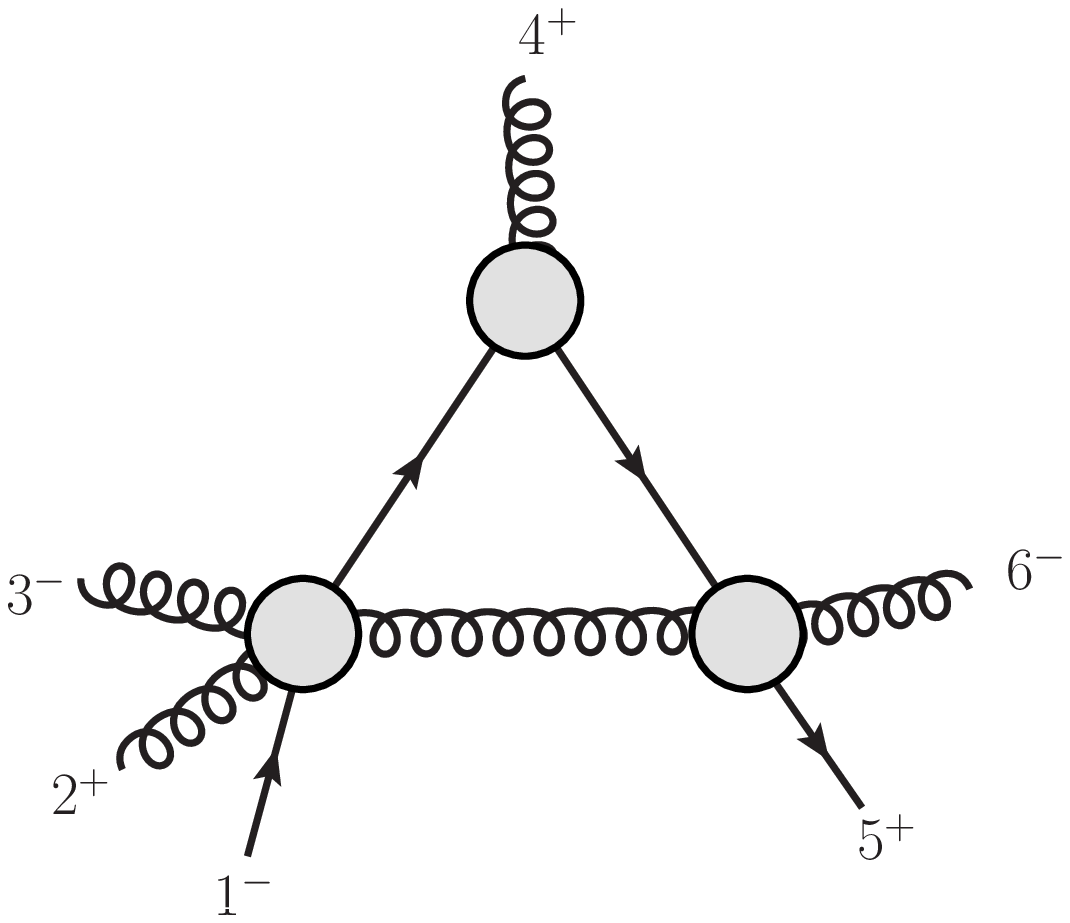}
  \end{minipage}
  \caption{Triple cut for two primitives with different helicity and ordering of external legs. The
  partial trees $A(\ell_{1;g}^{\pm},1_q^-,2_g^+,3_g^-,\ell_{4;\bar q}^{\pm})$ are the same for both
  primitives.}
  \label{fig:cache}
\end{figure}

An example is illustrated in Figure \ref{fig:cache} for a triangle cut of a six-point amplitude with
one external quark line and mixed loop content. Flipping the helicity of the fourth external gluon
and permuting the fifth and sixth external leg leaves the five point partial tree amplitudes
$A(\ell_{1;g}^{\pm},1_q^-,2_g^+,3_g^-,\ell_{4;\bar q}^{\pm})$ unchanged.

The relation between external permutation and loop momentum requires some comments. Let
$\tau\in\{2,3,4,5\}$ denote the topology of the cut, say bubbles, triangles, boxes and pentagons,
and ${\cal{K}}_\tau=\{k_1,...,k_{\tau-1}\}$ a set of $\tau-1$ external momentum sums which determine
the kinematic invariants of the cut. Since
the on-shell loop momentum for a distinct topology depends on  $\tau$  and on the
full set ${\cal{K}}_\tau$. A necessary condition to reuse loop momenta, is to find equal sets
of external momentum sums ${\cal{K}}_\tau$. In the above example, ${\cal{K}}_\tau$ is the same for
both permutations $\{\{1,2,3\},\{4\},\{5,6\}\}=\{\{1,2,3\},\{4\},\{6,5\}\}$. Analogously all six
permutations of the first three legs would again lead to the same loop
momenta. 
While the agreement of the external momenta and the flavour and helicity structure of
all legs is required directly by the partial tree amplitude itself, the loop momentum matching
depends on the topology and the permutation of all external legs. Hence, the knowledge of the cuts,
permutations, flavours and helicities allows us to assign every partial tree amplitude a unique
number which we use in a binary tree to cache those trees efficiently.

Since one has to sum over the helicities of the partons circulating in
the loop to determine a specific cut, we store always a block of
four amplitudes belonging to the four helicity combinations $(+,+)$,
$(+,-)$, $(-,+)$ and $(-,-)$ of the two partons connecting the
amplitude to the loop.

\subsection{Universal Pole Structure and Scheme Dependence}\label{sec:poles}

All basic quantities in \ngluon are computed in the four-dimensional
helicity (FDH) scheme. For consistency checks and the conversion to
the conventional dimensional regularisation (CDR) scheme the universal
structure of  infrared (IR) and ultraviolet (UV) poles is
required. The universal pole structure is known and can be derived from Refs. \cite{Catani:1996jh,Catani:2000ef}.
For the colour summed squared amplitudes in massless QCD the result 
in the FDH scheme reads:
\begin{align}
  &\operatorname{Re}(\mathcal{A}^{(0)\dagger}\cdot \mathcal{A}^{(1),FDH}_U) =
  \frac{1}{\eps^2}\left( N_c n_g + \frac{N_c^2-1}{2 N_c} n_q
  \right)g_s^2 | \mathcal{A}^{(0)}|^2 \nonumber\\&
  +\sum_{i=1}^{n-1} \sum_{j=i+1}^{n} \frac{1}{\eps} \log\left( \frac{\mu_R^2}{-s_{ij}} \right)g_s^2|\mathcal{A}^{(0)}_{ij}|^2
  -\frac{1}{\eps}\bigg( \beta_0
  +n_q \left( \frac{\beta_0}{2}-\frac{2}{3} \frac{N_c^2-1}{2N_c} \right)\bigg) g_s^2| \mathcal{A}^{(0)}|^2 \nonumber\\&
  + \text{ finite terms,}
  \label{eq:poles}
\end{align}
where $\mathcal{A}^{(1),FDH}_U$ denotes the unrenormalized one-loop amplitude in
the FDH scheme, $n_q$ is the number of external quarks and $n_g$ is the number of
external gluons and $g_s$ is the QCD coupling constant. The first coefficient of the QCD beta function is
given by $\beta_0 = \tfrac{11N_c-2 N_f}{3}$. The quantity
$|\mathcal{A}^{(0)}_{ij}|^2$ is the colour correlated Born amplitude squared defined by
\begin{align}
  |\mathcal{A}^{(0)}_{ij}|^2 = \langle \mathcal{A}^{(0)\dagger}| T_i \, T_j | \mathcal{A}^{(0)}\rangle
  \label{eq:born_cc}
\end{align}
where $T_g=f^{abc}$ and $T_q=T^a_{ij}$ and we followed the notation
used in \Ref{Catani:1996jh} to denote by $| \mathcal{A}^{(0)}\rangle$
the leading order amplitude which can be understood as a vector in
colour space. 
In order
to provide results in the conventional dimensional regularisation (CDR) scheme we apply the
conversion as determined by the universal IR and UV pole structure \cite{Catani:1996jh,Catani:2000ef}.
The scheme conversion to CDR and the renormalization are both
proportional to the Born amplitude. For convenience we list the explicit relations between
different prescriptions:
\begin{align}
  \frac{\operatorname{Re}\left(\mathcal{A}^{(0)\dagger}\cdot(\mathcal{A}^{(1),FDH}_{R}-\mathcal{A}^{(1),FDH}_{U})\right)}{g_s^2|\mathcal{A}^{(0)}|^2} &=
  -\frac{(n_g+n_q-2)}{2} \left( \frac{\beta_0}{\eps} - \frac{N_c}{3} \right) \\
  \frac{\operatorname{Re}\left(\mathcal{A}^{(0)\dagger}\cdot(\mathcal{A}^{(1),CDR}_{U}-\mathcal{A}^{(1),FDH}_{U})\right)}{g_s^2|\mathcal{A}^{(0)}|^2} &=
  \frac{(n_g+n_q-2)}{2}\frac{N_c}{3} - \frac{n_g N_c}{6} - \frac{n_q}{4 N_c} \left( N_c^2-1 \right) \\
  \frac{\operatorname{Re}\left(\mathcal{A}^{(0)\dagger}\cdot(\mathcal{A}^{(1),CDR}_{R}-\mathcal{A}^{(1),FDH}_{U})\right)}{g_s^2|\mathcal{A}^{(0)}|^2} &=
  -\frac{(n_g+n_q-2)}{2} \left( \frac{\beta_0}{\eps} - \frac{N_c}{3} \right)\nonumber\\
  &\phantom{{}={}}{}-\frac{n_g N_c}{6} - \frac{n_q}{4 N_c} \left( N_c^2-1 \right)
  \label{eq:re}
\end{align}
where the subscript $R$ denotes renormalized amplitudes. 

\section{Installation and Usage \label{sec:install}}

The code can be obtained from \url{https://bitbucket.org/njet/njet/} as a tar archive.
Alternatively the source can be obtained from the public git repository using
\begin{verbatim}
  git clone https://bitbucket.org/njet/njet.git
\end{verbatim}
Compilation for a wide variety of architectures and compilers is
supported via a configure script generated with autoconf. The default
set-up will be configured by simply using
\begin{verbatim}
  ./configure
\end{verbatim}
which will scan your system for the required pre-requisites. 
Installation into a local directory structure is supported through the option \verb~--prefix=<dir>~. 
Further useful options supported by the configure script are:
\begin{itemize}
  \item \verb1--enable-demo1 Will compile some additional example
    programs which can be run with \verb1make check1
    after the library has been successfully compiled. These demo programs illustrate the usage of the program
    and the correctness of the implementation. A sample output is
    given in appendix \ref{app:numchecks}.
  \item \verb1--enable-5jet1 Includes in the library the one-loop corrections  for
    five jet production. Since the compilation requires a significant
    amount of memory the one-loop corrections for five jet production are disabled by
    default. It is not recommended to use this option with less then 4~GB of RAM.
  \item \verb1--disable-cache1 Per default the cache system is enabled. It allows caching of partial tree amplitudes to
    optimise the colour and helicity sums as described in section
    \ref{sec:cache}. Due to the way the cache is implemented
    an architecture dependent restriction is imposed on the total
    number of external particles. This is 6 on a 32-bit system and 13 on a 64-bit system.
    By using \verb~--disable-cache~ the cache can be disabled and the
    restrictions can be lifted however at the cost of losing some speed.
\end{itemize}
A complete list of possible options can be obtained using \verb1--help1.
Further details can be found in the INSTALL file in the root directory of the distribution.
Once the configure script has successfully generated the makefiles compilation and installation
can be performed with:
\begin{verbatim}
  make
  make install
\end{verbatim}
Additionally,
\begin{verbatim}
  make check
\end{verbatim}
will run some basic tests to verify a correct installation.
\NJet is distributed with a version of
\texttt{QCDLoop} for the evaluation of the scalar one-loop integrals
and \texttt{qd}~\cite{Bailey:QD} to extend the precision of the
floating-point arithmetic to quadruple precision.

\subsection{Interface via the Binoth Les Houches Accord}

Though the library can be interfaced directly, the simplest way to use it is to link
via the Binoth Les Houches Accord (BLHA) interface~\cite{Binoth:2010xt}. Once correctly installed, a python
script \verb1njet.py1 will interpret a user supplied ``order file" to produce a ``contract file"
allowing the various processes and options to be passed to \NJet.
Several sample programs demonstrating the usage of this interface are included in the distribution with
corresponding order files in the sub-directory \verb~examples~. These
examples can be compiled independently using the provided sample
Makefile. In the following we briefly illustrate an example.
The order file used in connection with the  BLHA interface consists of two parts.
The first part contains general settings such as matrix element type and regularisation scheme.
The second part is a list of elementary sub-processes specified using PDG particle codes~\cite{Yost:1988ke}.
An example file for $u \ubar \to \text{3-jets}$ is shown below:
\begin{framed}
\begin{verbatim}
# order file for u ubar -> jets
MatrixElementSquareType CHsummed
CorrectionType          QCD
IRregularisation        CDR
AlphasPower             3
# process list
1 -1 -> 21 1 -1
1 -1 -> 21 2 -2
1 -1 -> 21 21 21
\end{verbatim}
\end{framed}
Assuming the order file is saved with the file name \verb~OLE_order.lh~ the command
\begin{verbatim}
  njet.py -o OLE_contract.lh OLE_order.lh
\end{verbatim}
will process it and creates a contract file \verb~OLE_contract.lh~.
The user should always check the generated contract file to ensure that all
requested parameters and sub-processes have been recognized by \NJet
which is indicated in the generated contract file.
Having a valid contract file one can use it with the standard BLHA interface functions
which are declared in the \verb~njet.h~ header:
\begin{verbatim}
void OLP_Start(const char* filename, int* status);
void OLP_EvalSubProcess(int mcn, double* pp, double mur,
                        double* alphas, double* rval);
\end{verbatim}
The conventions of their usage are described in \cite{Binoth:2010xt}, we outline
here only the main features. Firstly the interface has to be initialized
by calling the function \verb1OLP_Start1, passing the file name of a contract file as a parameter.
After the call, the second parameter \verb1status1 will contain \verb~1~ indicating the success or
\verb~0~ indicating the failure of the initialization.
In case of a successful initialization one can begin using
\verb1OLP_EvalSubProcess1 to evaluate the subprocess \verb~mcn~ for a given
phase space point with momenta \verb1pp1, renormalization scale \verb1mur1 and
strong coupling \verb1alphas1. The integer \verb~mcn~ is the number of the process in the
contract file (starting from~1). The momenta are passed through the one
dimensional array \verb1pp1. Note that in addition to the momentum for each
parton also its mass is transferred, e.g.
\begin{equation}
  \verb1pp[5n]1 = \{p^{0}_0, p^{1}_0, p^{2}_0, p^{3}_0, m_0,p^{0}_1,
  p^{1}_1, p^{2}_1, p^{3}_1, m_1,\ldots,p^{0}_{n-1}, p^{1}_{n-1}, p^{2}_{n-1}, p^{3}_{n-1},m_{n-1}
  \},
\end{equation}
where $p_n^i$ denotes the $i$-th momentum component of the $n$-th parton.
The result is returned through the array \verb1rval1 with the
following order of elements: $1/\eps^2$-pole, $1/\eps$-pole, finite part, tree amplitude.
If requested in the order file, the \verb1rval1 array will contain three additional elements:
relative accuracy of the $1/\eps^2$-pole, relative accuracy of the
$1/\eps$-pole, and
relative accuracy of the finite part.

\subsubsection{Additional BLHA options}

In addition to the standard options (see Ref.~\cite{Binoth:2010xt}) \verb1njet.py1 allows the following additional
options to control the interface.
\begin{itemize}
  \item \verb1NJetType [default = loop]1\\
    \verb1loop1, \verb1loopds1 or \verb1tree1 may be specified to
    determine the type of the squared matrix element  returned.
    \verb1loop1 and \verb1tree1 are conventional colour summed squared
    amplitudes for one-loop and tree contributions respectively.
    \verb1loopds1 returns the de-symmetrized sums if available otherwise the standard one-loop sum
    is returned. A more advanced usage is to specify a list separated by a `-' of a process number
    followed by a `.' and the type required for that process (see example in supplied demo program \verb~examples/DSYMtest~).
  \item \verb1NJetReturnAccuracy [default = no]1\\
    If set to \verb1yes1, the scaling test is used to estimate the
    accuracy and the result is returned in the output array making it three entries longer.
  \item \verb9NJetSwitchAcc [default = 1e-5]9\\
    Used to set the relative accuracy at which higher precision floating point
    arithmetics will be used to evaluate the phase space point if double
    precision has not produced the desired accuracy.
    To account for the statistical nature of the scaling test it is recommended to specify 1 or 2 digits
    more than the required accuracy.
  \item \verb~NJetNf [default = 5]~\\
    Set the number of light quark flavours circulating in the fermion loops.
\end{itemize}
A full documentation of the usage of \verb1njet.py1 can be found in \verb1lh/LH.txt1.

\subsection{The {\tt NParton} library for multi-parton primitives}

Also provided with the package is the \verb1NParton1 class for the evaluation of arbitrary
multi-parton primitive amplitudes both at tree-level and at one loop. The usage is demonstrated in
an example code compiled into the \verb1src-nparton1 directory: \verb1NParton-test.cpp1.

\subsubsection{Initialization and Namespaces}

\verb1NParton1 is designed such that both double and quadruple
precision can be used in parallel in the same
program. In order to achieve this, different namespaces \verb1NJet_sd1 and \verb1NJet_dd1 are used
for single double and double double (=quadruple) respectively. Inclusion of the header file {\tt
src-nparton/NParton-multiprec.h} will take care of the necessary declarations.

Both classes require loop integrals and other parameters to be initialized before they can be
used which is done by creating the static objects
\begin{verbatim}
static NJet_sd::Initialize global_init_sd;
static NJet_dd::Initialize global_init_dd;
\end{verbatim}
before the start of the main program.

\subsubsection{Constructor}
The calculation of the primitive amplitudes is hidden in the class \verb1NParton1.
The class constructor only requires one integer argument specifying the number of external legs.
The namespace for the required accuracy must also be specified, e.g.
\begin{verbatim}
NJet_sd::NParton primitive(int npartons);
\end{verbatim}
for double precision and,
\begin{verbatim}
NJet_dd::NParton primitive(int npartons);
\end{verbatim}
for quadruple precision.

\subsubsection{Member functions}
The actual calculation is controlled through member functions of the
\verb1NParton1 class. The relevant functions are:
\begin{itemize}
  \item \verb1void setAmp(double moms[][4], int[] flavs)1 \\
   This function sets the momenta and flavour of the external legs being the same for all helicity
   configurations of one primitive amplitude.  The four-momenta are entered as an array of
   four-vectors with the energy in the zeroth component according to $(E,p_x,p_y,p_z)$. The list of
   flavours is specified as follows: 0 for a gluon, $(-)i$ for each
   (anti-)quark with $i\in \{1,2,...,{{N}}_{q\bar q}\}$ and ${{N}}_{q\bar q}$ the total number of
   external quark lines. Due to fermion and flavour number conservation the
   quarks must always appear as quark--anti-quark pairs. 
   Each quark line must be labelled with a different integer. The
   primitive amplitudes calculated thus correspond to the case that
   all quark flavours are different. The equal flavour case can be
   obtained from these results through an appropriate anti-symmetrization. 
   Since QCD conserves the flavour number the fermion lines are not
   allowed to cross in the primitive amplitudes. A valid choice would be, for example,
   \verb9{1,0,2,-2,-1}9 while \verb9{1,0,2,-1,-2}9 would however give an error.
 \item \verb1void eval(int[] hels, int ptype)1 \\
   \verb1void eval(int hels, int ptype)1 \\
   After a previous call of \verb1setAmp1 this function evaluates a
   single primitive amplitude. The first argument is a list of
   helicities specified by $\pm1$ for each parton. Alternatively the
   helicity can be given as a single integer between 0 and $2^n-1$
   which is very convenient to loop over. The second argument selects
   either the mixed or the fermion loop primitive amplitude using the
   internally defined constants \verb1NParton::mixed1 and
   \verb1NParton::qloop1.
 \item \verb1std::complex<double> evalTree(int[] hels)1 \\
   \verb1std::complex<double> evalTree(int hels)1 \\
   Similar to the \verb1eval1 function except that only the tree
   amplitude is evaluated. As \verb1eval1 this function requires a
   previous call of {\tt setAmp}. The value of the tree-level helicity
   amplitude is returned.
  \item \verb1std::complex<double> getAmp(int p)1 \\
   \verb1std::complex<double> getAbsErr(int p)1 \\
   \verb1std::complex<double> getRelErr(int p)1 \\
   After calling the \verb1eval1 function the various components of the amplitudes and their absolute
   and relative numerical uncertainty can be accessed with these
   functions. The different contributions are obtained by setting
   the variable \verb1p1 to one of the predefined constants:\\
\begin{tabular}{ll}
 {\tt NParton::eps0} & returns the finite part, \\
 {\tt NParton::eps1} & returns the coefficient of the $1/\epsilon$ pole, \\
 {\tt NParton::eps2} & returns the coefficient of the $1/\epsilon^2$ pole, \\
 {\tt NParton::tree} & returns the tree-level amplitude, \\
 {\tt NParton::rational} & returns the rational part, \\
 {\tt NParton::cutconstructible} & returns the cut-constructible part.\\
\end{tabular}
  \item \verb1void IRcheck()1\\
    After calling the \verb1eval1 function this function will check the poles against known
    universal pole structure. The relative error of both $1/\eps^2$ and $1/\eps$ poles compared with
    the analytic formulae will be displayed.
  \item \verb1void refine(double moms, dd_real moms_dd)1 \\
    This function  converts a double precision phase-space point to
    quadruple precision to enable the calculation in quadruple
    precision using \verb1NJet_dd::NParton1. 
  \item \verb1void setMuR2(double MuR2)1 \\
    Used to set the renormalization scale squared, $\mu_R^2$.
  \item \verb1std::string ProcessStr()1 \\
    Returns a string with the current flavour and helicity structure. This gives the state
    of the last \verb1evalAmp1 call.
\end{itemize}
With the exception of the \verb1refine1 function, when using \verb1NJet_dd::NParton1, \verb1double1 should be
replaced with \verb1dd_real1.

\section{Results \label{sec:results}}

\subsection{Validation}

\begin{table}
  \centering
  \begin{tabular}{lc}
    Process & Reference \\
    $pp+2j$ (all $2\to2$ with colour summation) & \cite{Ellis:1985er} \\
    $5g$ & \cite{Bern:1993mq} \\
    $q\qbar +3g$ & \cite{Bern:1994fz} \\
    $n(g)$ (finite) & \cite{Bern:2005hs} \\
    $n(g)$ (MHV) & \cite{Forde:2005hh} \\
    $q\qbar+n(g)$ (finite) & \cite{Bern:2005ji}
  \end{tabular}
  \caption{Cross checks against analytic expressions known in the literature.}
  \label{tab:analyticchecks}
\end{table}

To assess the correctness of our implementation we performed a series
of highly non-trivial tests. The first test is to check the universal
structure of the IR and UV poles (see section \ref{sec:poles}). Both
at the level of primitive amplitudes and colour summed amplitudes we
find for all the processes we have tested perfect agreement for the
$1/\epsilon^2$ and $1/\epsilon$ poles where $D=4-2\epsilon$ with $D$
denoting the dimension of space-time. As far as the UV poles are
concerned this check tests a linear combination of the coefficients of
the two-point scalar integrals. The correct IR structure is sensitive
to the coefficients of all IR divergent scalar three- and four-point
integrals. Since the coefficients of the scalar integrals are all
calculated the same way reproducing the correct IR and UV structure
provides a strong test on the entire cut-constructible
part.\footnote{We note that a possible small deviation of the
  numerically computed poles from the predicted pole structure due to
  numerical rounding errors could give some indication of the
  numerical accuracy of the current evaluation.}  However the rational
part of the amplitude cannot be tested in this way.  To verify the
non-trivial rational part a large number of cross checks against known
processes at individual phase-space points has been performed. We have
confirmed the results for primitive amplitudes for specific helicity
configurations with the known analytic results listed in table
\ref{tab:analyticchecks}. The checks of the primitive amplitudes in
pure gauge theory with the all $n$ multiplicity formulae in
\cite{Bern:2005hs,Forde:2005hh} have already been presented in
\Ref{Badger:2010nx}. Primitive amplitudes with one external quark pair
have been tested against the explicit formulae for $q \qbar+3g$ given
in \Ref{Bern:1994fz}. We find complete agreement for all independent
primitive amplitudes and all independent helicity configurations for
these five-point amplitudes. The IR finite amplitudes with a single
quark pair, $A_n^{[m/f]}( q_1^-,g_2^+,...,\qbar_i^+,...g_n^+)$, have
been computed analytically in \Ref{Bern:2005ji} to all multiplicity.
Direct comparisons over all possible configurations of the quark pair
have been performed and in all cases we find good numerical agreement.
The program \verb1NParton-qqchecks1 in the directory \verb1src-tools1
performs the comparisons with the analytic formulae listed above.
\begin{table}[h]
  \begin{center}
    \begin{tabular}{|c|c|c|}
      \hline
      & \NJet & \BlackHat \\
      \hline
      \multicolumn{3}{|c|}{$g g \to g g g g$} \\
      \hline
      born       & $+4.9769357371794838\times 10^{8}$ & $+4.976935736\times 10^{8}$ \\
      1/$\eps^2$ & $-1.7999999999999108\times 10^{1}$ & $-1.800000000\times 10^{1}$ \\
      1/$\eps$   & $-6.5144862057710100\times 10^{1}$ & $-6.514486205\times 10^{1}$ \\
      finite     & $-3.2130366101334992\times 10^{1}$ & $-3.213036625\times 10^{1}$ \\
      \hline
      \multicolumn{3}{|c|}{$d \dbar \to g g g g$} \\
      \hline
      born       & $+2.1622011190045194\times 10^{5}$ & $+2.162201118\times 10^{5}$ \\
      1/$\eps^2$ & $-1.4666666666663076\times 10^{1}$ & $-1.466666667\times 10^{1}$ \\
      1/$\eps$   & $-5.8264471151950865\times 10^{1}$ & $-5.826447114\times 10^{1}$ \\
      finite     & $-4.3957884552089730\times 10^{1}$ & $-4.395788455\times 10^{1}$ \\
      \hline
      \multicolumn{3}{|c|}{$d \dbar \to u \ubar g g$} \\
      \hline
      born       & $+1.3745823177248822\times 10^{4}$ & $+1.374582317\times 10^{4}$ \\
      1/$\eps^2$ & $-1.1333333333333261\times 10^{1}$ & $-1.133333333\times 10^{1}$ \\
      1/$\eps$   & $-4.8019618061344993\times 10^{1}$ & $-4.801961805\times 10^{1}$ \\
      finite     & $-3.7348157184728159\times 10^{1}$ & $-3.734815718\times 10^{1}$ \\
      \hline
      \multicolumn{3}{|c|}{$d \dbar \to u \ubar s \sbar$} \\
      \hline
      born       & $+3.6761219414819656\times 10^{1}$ &  $+3.676121941\times 10^{1}$ \\
      1/$\eps^2$ & $-8.0000000000001386$ &               $-8.000000000$ \\
      1/$\eps$   & $-2.6337813992821804\times 10^{1}$ &  $-2.633781399\times 10^{1}$ \\
      finite     & $-6.7242846898320696\times 10^{-1}$ & $-6.724284689\times 10^{-1}$ \\
      \hline
    \end{tabular}
  \end{center}
\caption{Numerical comparison between \NJet and the $2\to4$ amplitudes given in Ref.
\cite{Bern:2011ep}.}
\label{tab:BHcompare}
\end{table}
In addition---with the exception of the six-gluon channel---we have cross checked numerically the colour summed
interferences for all processes contributing to three and four jet
production with results generated with {\sc HELAC-1LOOP}~\cite{Bevilacqua:2011xh}. 
We were also able to confirm results for the fermion loops with those generated
from {\sc GoSam} \cite{Cullen:2011ac}. 
These numerical
checks against existing codes can be accessed through the sample
program \verb1NJet-demo1. Sample output is provided
in appendix \ref{app:numchecks}.
In case of four jet production full agreement has also been
found with results for a single phase-space point provided by \BlackHat \cite{Bern:2011ep}. 
The explicit outcome is shown in Table~\ref{tab:BHcompare}. 
The results in \Ref{Bern:2011ep} are given up to ten decimal digits. Apart from the gluonic channel
we recover all digits using double precision. For the pure gluon
channel the error estimate of the numerical evaluation obtained via the
scaling test gives,
\[ A^{(1),\text{finite}}_{6g} = -3.2130366101334992\times 10^{1} \pm 3\times 10^{-8}, \]
hence all reliable digits are in agreement with the \BlackHat result. When evaluated in quadruple
precision all digits are in agreement:
\[ A^{(1),\text{finite}}_{6g} = -3.2130366250275191\times 10^{1} \pm 2\times 10^{-13}. \]
The full set of quadruple precision
checks can be reproduced using \verb1NJet-demo-dd --BlackHat4j1.

It is important to note that the scalar one-loop integrals are evaluated in double precision which
limits the final accuracy of \NJet to 16 decimal digits even when quadruple precision is used. 

Given the large number of checks passed by \NJet we are confident the
library produces reliable results for one-loop amplitudes in
massless QCD.

Since the amplitudes for virtual corrections to five-jet production have not been presented
in the literature before, we give the numerical values for the evaluation at a single phase space
point in appendix \ref{app:7parton}.

\subsection{Accuracy \label{sec:acc}}

Given the complexity of the calculations performed in the \NJet
library it is important to have a precise understanding of the
numerical accuracy. We have performed a number of tests on
the output of both primitive amplitudes and colour summed squared
matrix elements. By performing the scaling test as described in
\Ref{Badger:2010nx} we can reliably estimate the number of correct digits returned by the program.

\begin{figure}[h]
  \begin{center}
    \begin{tabular}[h]{cc}
      \includegraphics[width=7cm]{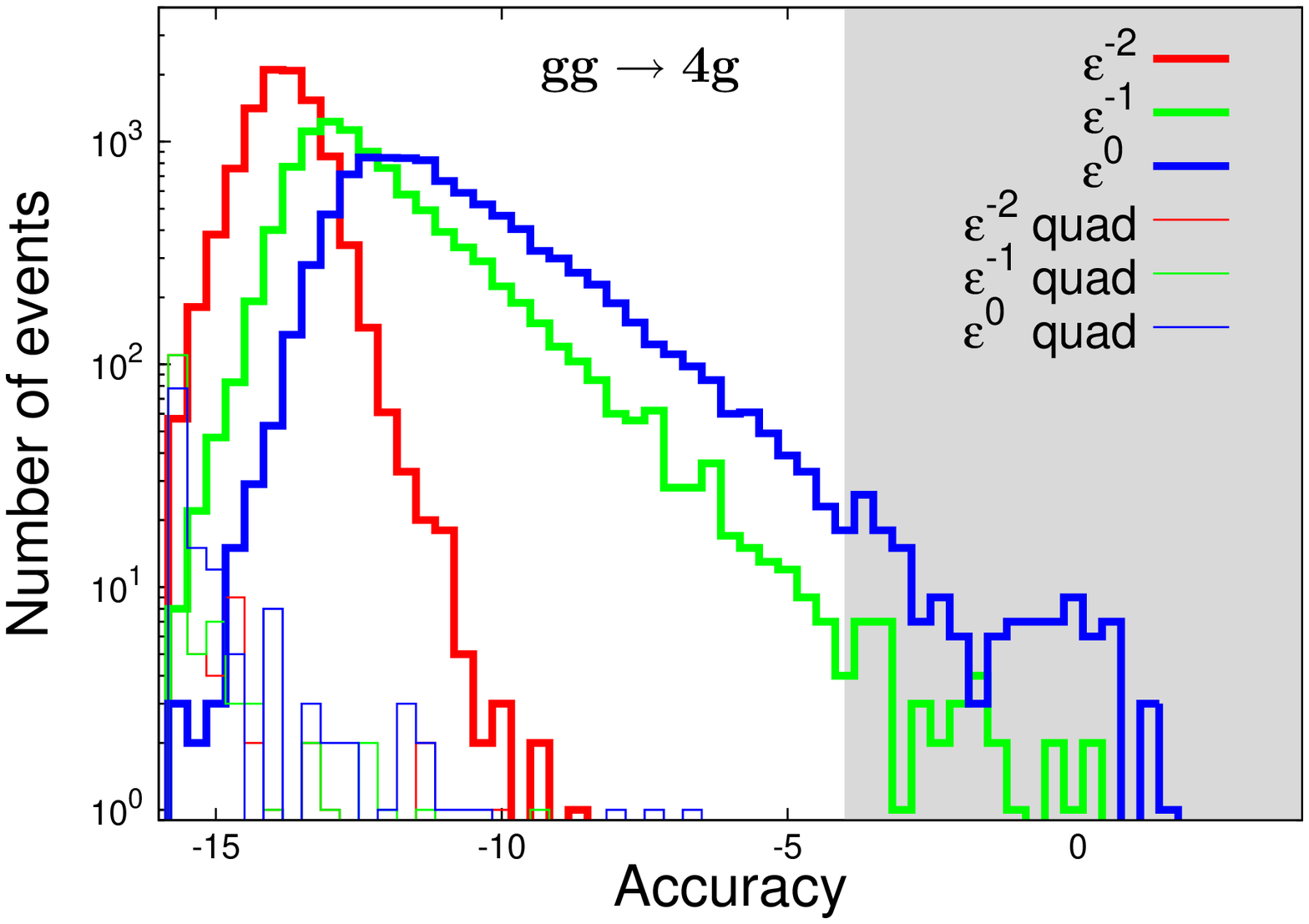} &
      \includegraphics[width=7cm]{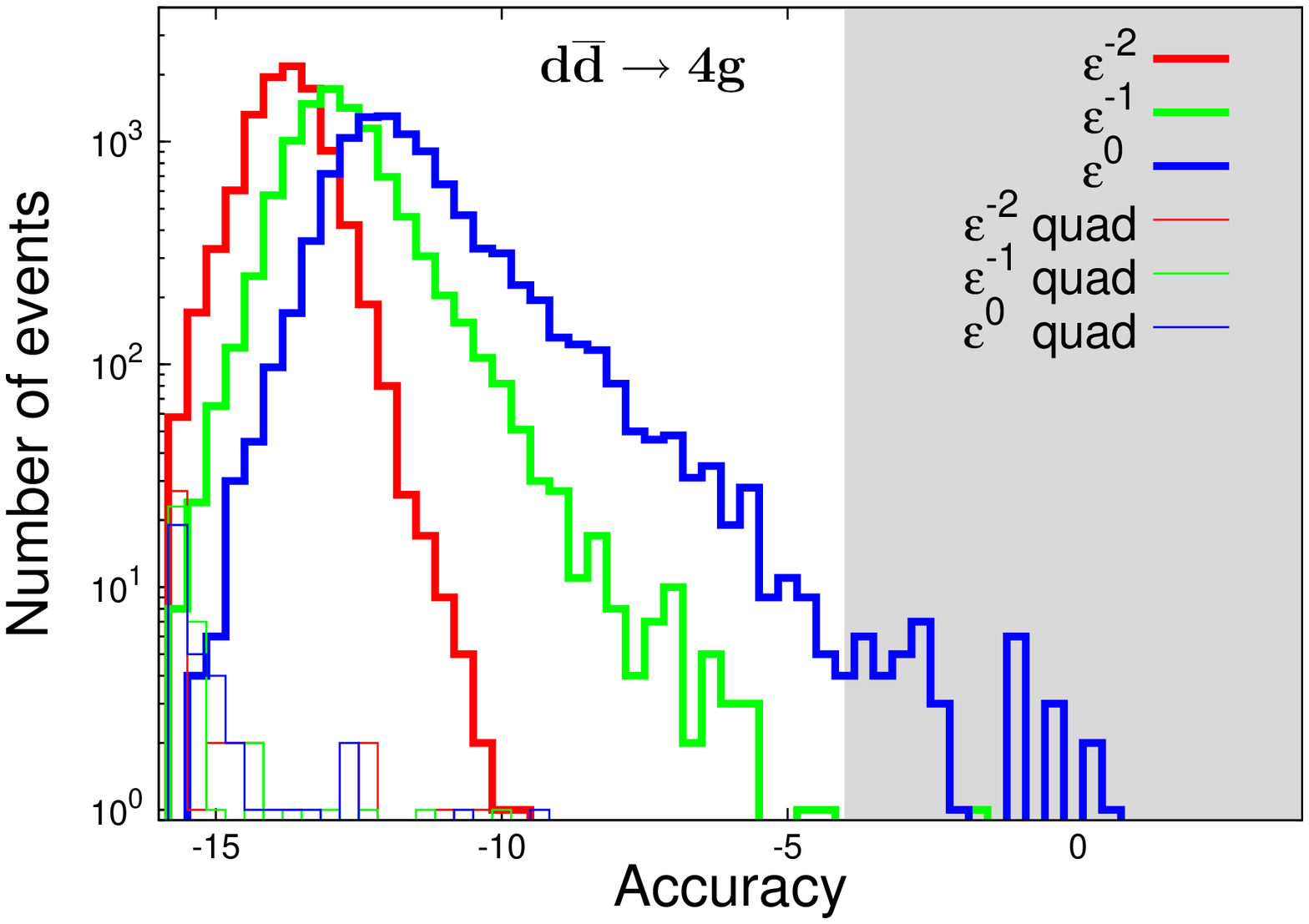} \\
      (a) & (b) \\
      \includegraphics[width=7cm]{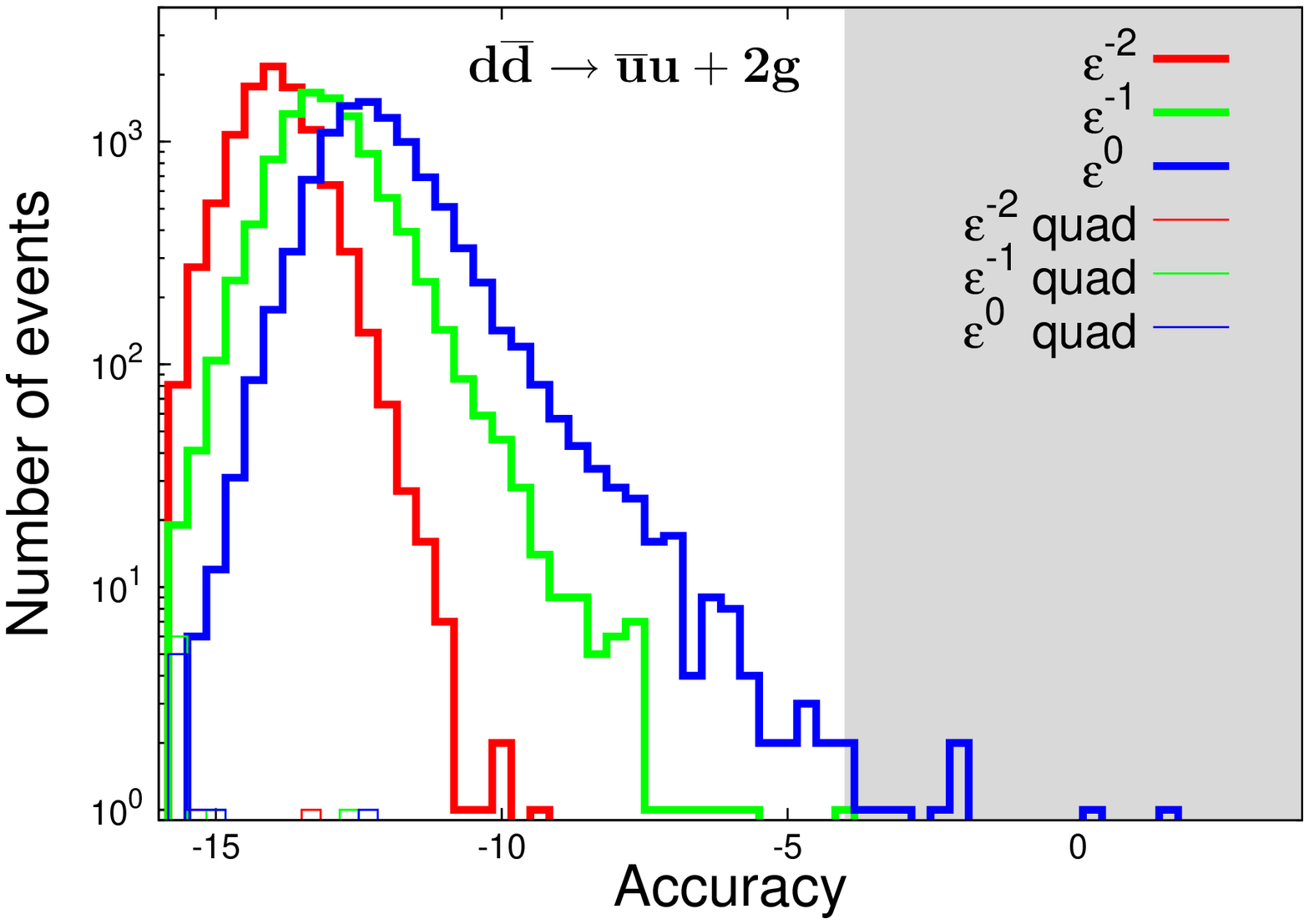} &
      \includegraphics[width=7cm]{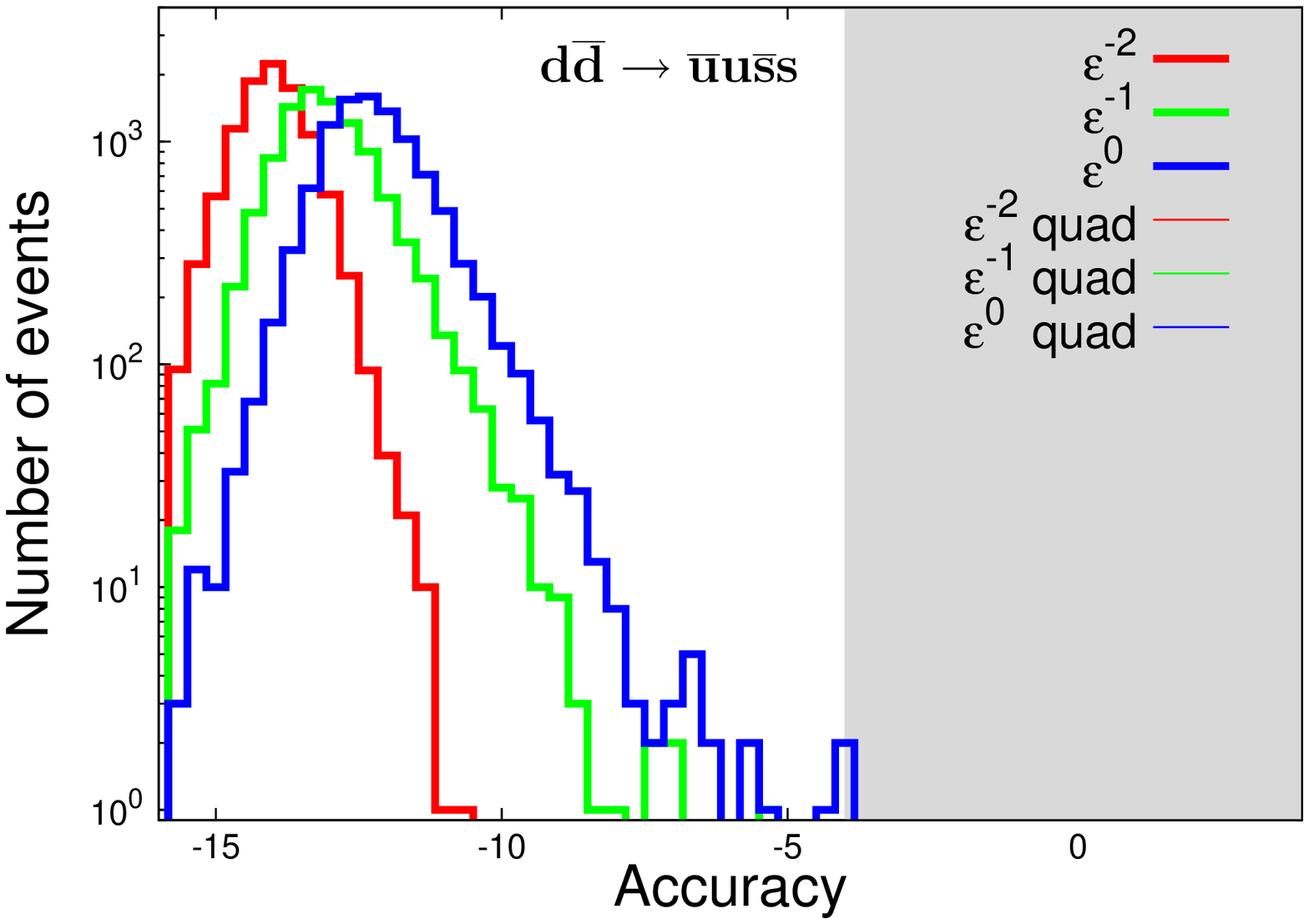} \\
      (c) & (d)
    \end{tabular}
  \end{center}
  \caption{Accuracy for 4-jet amplitudes: (a) shows the six gluon process, (b) the $d\dbar \to 4g$
  process, (c) $d \dbar \to \ubar u + 2g$ and (d) the six unlike quark pair process.  The thicker
  histograms show computations in double precision whereas the thinner curves show the distribution
  in quadruple precision for points which did not pass the relative
  accuracy of $10^{-4}$ when calculated in double precision (region 
  marked by the shaded area). Red histograms show the $\tfrac{1}{\eps^2}$ poles, green histograms the
  $\tfrac{1}{\eps}$ and blue histograms the finite part of the amplitudes.}
  \label{fig:acc_4j}
\end{figure}

Figure \ref{fig:acc_4j} shows for four channels contributing
to four jet production the  distribution
of the estimated accuracy. In total $10^4$ phase space points were
generated. The accuracy is defined by
\begin{eqnarray}
  \text{Accuracy}(\mathcal{A}(p,\mu_R)) =
  \log_{10}\left(2\frac{\mathcal{A}(p,\mu_R)-x^{n-4}\mathcal{A}(xp,x\mu_R)}{\mathcal{A}(p,\mu_R)+x^{n-4}\mathcal{A}(xp,x\mu_R)}\right),
\end{eqnarray}
where $p$ are the external momenta, $\mu_R$ is the renormalization and $x$ is the scaling parameter.
This definition gives a direct measure of the valid digits.
The phase space points were generated with a
sequential splitting algorithm as described in
\cite{Byckling:1973}. For large multiplicities the algorithm prefers
collinear configuration. Compared to a flat phase space generation we
believe that this simulates better what will actually happen in a
phase space integration.
A weak phase space cut of $|p_i\cdot p_j| > s\times10^{-2}$ was made for a
centre-of-mass energy of $s=7$ TeV. For the most complex channels around $1\%$ of points must be
re-evaluated in quadruple precision. Some of them are related to small Gram determinants but these
are not the only sources of numerical instability within the algorithm.  As we consider channels
with more quarks, and therefore lower maximum tensor rank, the
accuracy improves considerably. For these processes all evaluations pass the threshold accuracy of $10^{-4}$ after
quadruple precision has been used. Similar, though  significantly improved behaviour, is observed for
the lower multiplicity channels, plots of the distributions can be found in Ref.~%
\cite{Badger:2012dd}.

\begin{figure}[h]
  \begin{center}
    \begin{tabular}[h]{cc}
      \includegraphics[width=7cm]{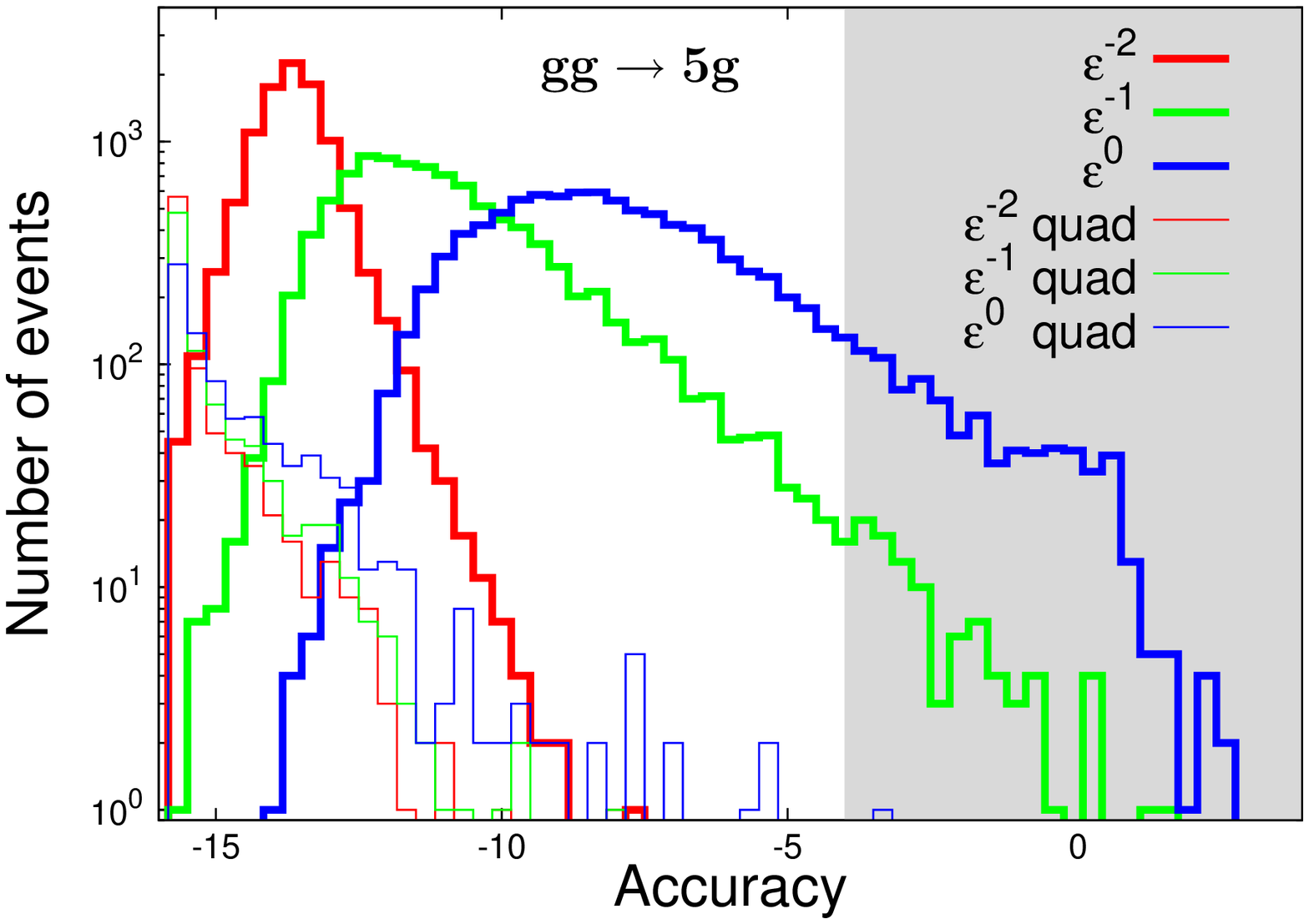} &
      \includegraphics[width=7cm]{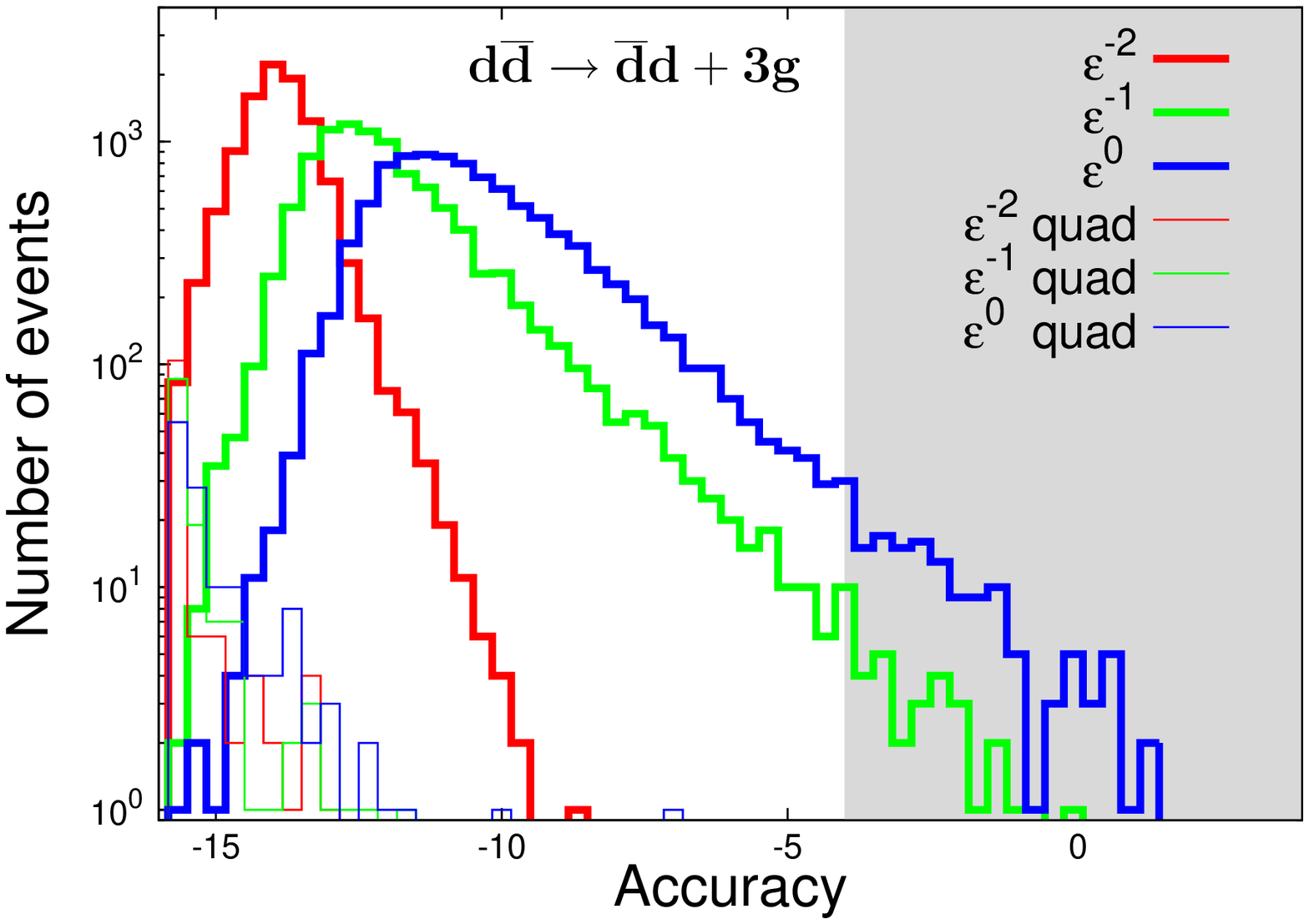} \\
      (a) & (b)
    \end{tabular}
  \end{center}
  \caption{Accuracy for 5-jet amplitudes: (a) shows the seven gluon process and (b) the $d \dbar \to \dbar d +
  3g$ process. The thicker histograms show computations in double precision whereas the thinner
  curves show the distribution in quadruple precision for points which did not pass the relative
  accuracy of $10^{-4}$ when calculated in double precision. Red histograms show the $\tfrac{1}{\eps^2}$
  poles, green histograms the  $\tfrac{1}{\eps}$ and blue histograms the finite part of the amplitudes.}
  \label{fig:acc_5j}
\end{figure}

Figure \ref{fig:acc_5j} shows an identical analysis for two channels contributing to five jet
production. For the seven gluon amplitude we observe an increasing number of instabilities such
that the time to re-evaluate with quadruple precision becomes noticeable. Again increasing the
number of fermion pairs improves the situation a lot. Performing the
floating point arithmetics in  quadruple precision $99.99\%$ of points
achieve the required 4~digits of accuracy. 
It might be required switching to octuple precision when performing phase-space integration of virtual
amplitudes of this complexity.

\subsection{Speed Tests}

\begin{figure}[t]
  \begin{center}
    \includegraphics[width=0.8\textwidth]{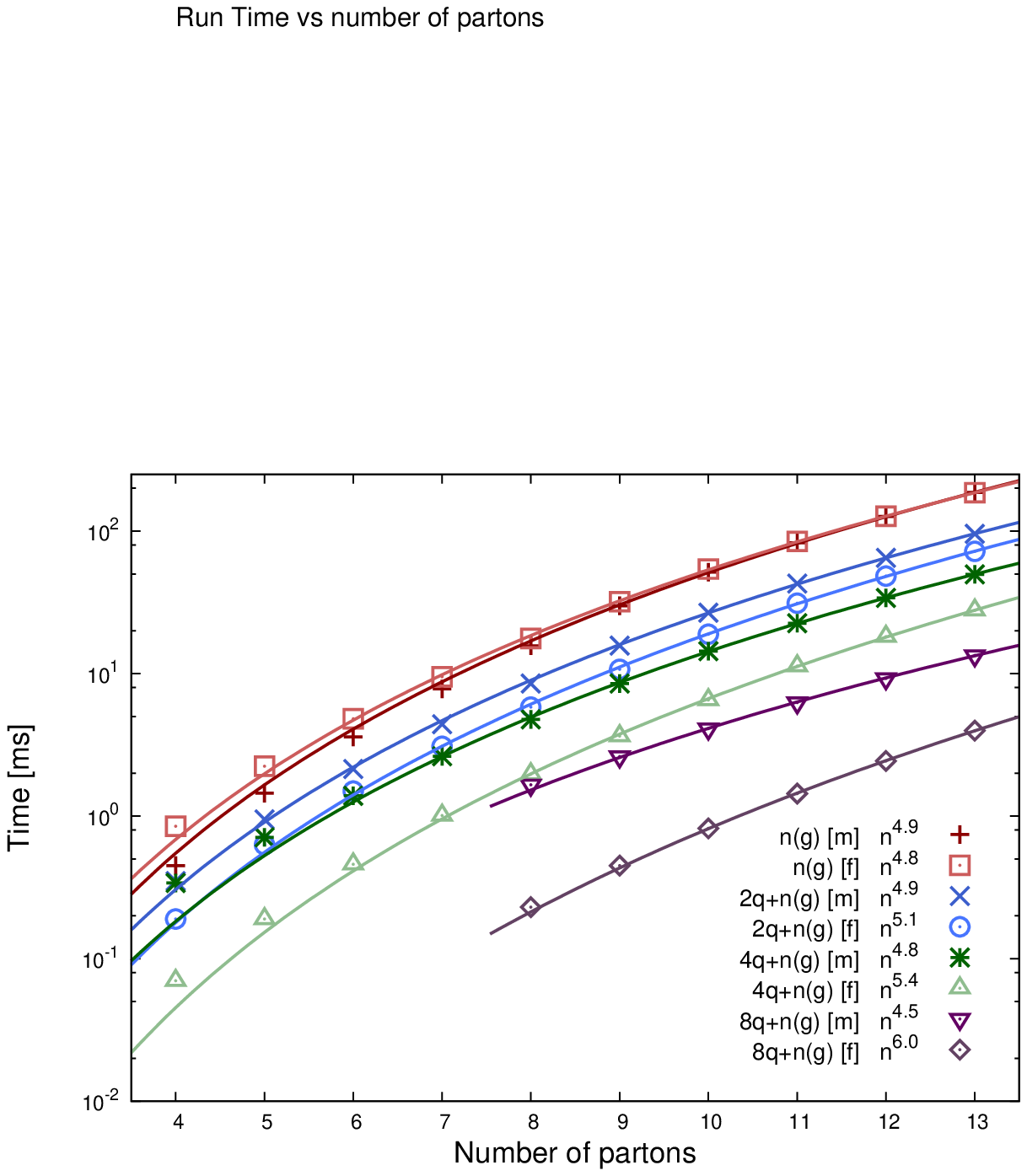}
  \end{center}
  \caption{Estimated evaluation time for primitive amplitude with 0, 1, 2 and 4 fermion lines
  averaged over the total number of helicities as a function of the number of external legs. Both
  mixed quark gluon loop and closed quark loop are shown. The curves are a fit to the polynomial
  $f(n)=an^b$. The exponent $b$ can be read off from the legend. The times were measured on an
  Intel(R) Core(TM)2 Duo CPU E8400 @ 3.00GHz.}
  \label{fig:primitivetimes}
\end{figure}

An advantage of the recursive construction of ordered tree-level amplitudes within \ngluon is that
the computation of multi-fermion primitives scales polynomially with time. The scaling properties of
the algorithm for up to three quark pairs with fixed parton ordering and up to 13~external legs,
the maximally allowed number with the cache system turned on, are shown in
Figure~\ref{fig:primitivetimes}.
For simplicity we restrict the analysis to amplitudes with neighbouring quark pairs, $A_n(q_1,\overline
q_1,...,q_k,\overline q_k,g_{2k+1},...,g_n)$ for $k$ quark lines. The times are averaged over all
$2^n$ helicity configurations including amplitudes that vanish trivially by helicity conservation
along fermion lines. The scaling behaviour reflects that the cache system has been used to improve
the efficiency of the helicity sum and varies between $n^{4.5}$ and $n^{6.0}$. In the asymptotic
region the algorithm employed in \ngluon is known to scale as $n^8$ for the mixed loop content
\cite{Badger:2010nx}. For the closed quark loop this is reduced to $n^7$ since the coupling of
external off-shell currents to the fermion loop goes exclusively via
three-point vertices. We find that
the asymptotic scaling behaviour is reached at extremely high multiplicities ($>20$ legs) and has
little practical relevance. The asymptotic scaling holds also for the average of configurations with
non-neighbouring quark pairs. Of course, the timings of individual amplitudes are sensitive to a
non-vanishing quark antiquark separation. One observes the expected pattern of a speed up in the
closed quark loop case and a slow down in the mixed quark gluon loop. A more detailed analysis for
non-neighbouring quark configurations is given in \cite{Badger:2011zv}.

Table \ref{tab:timechsum} shows the timing estimates for colour and helicity summed virtual corrections for all
sub-processes of multi-jet production with up to $5$ jets. The times are quoted for single double
precision, $T_{sd}$, and including corrections using quadruple precision in order to achieve an
overall accuracy of $10^{-4}$, $T_{4 \text{ digits}}$. We also show
the fraction of points which did not pass the relative accuracy when
evaluated in double precision. All
times include the two evaluations necessary to obtain the accuracy estimate via the scaling test.
Times have been obtained using the {\tt clang} v2.8 compiler. The time
estimated for the reevaluation using quadruple precision
is based on the percentage of rescued points from the accuracy tests in the previous section.
On average quadruple precision evaluation was found to be between 7 and 8 times longer than double precision.
We stress that even in the worst case of the seven gluon amplitude the
required reevaluation only doubles the total run time.
\begin{table}
  \begin{center}
  \begin{tabular}[h]{|lccc|}
    \hline
    process & $T_{sd}$[s] & $T_{4 \text{ digits}}$[s] & (\% fixed) \\
    \hline
    4g   & 0.030 & 0.030 &(0.00) \\
    2u2g & 0.032 & 0.032 &(0.00) \\
    2u2d & 0.011 & 0.011 &(0.00) \\
    4u   & 0.022 & 0.022 &(0.00) \\
    \hline
  \end{tabular}
  \begin{tabular}[h]{|lccc|}
    \hline
    process & $T_{sd}$[s] & $T_{4 \text{ digits}}$[s] & (\% fixed) \\
    \hline
    5g     & 0.22 & 0.22 &(0.22) \\
    2u3g   & 0.34 & 0.35 &(0.06) \\
    2u2d1g & 0.11 & 0.11 &(0.00) \\
    4u1g   & 0.22 & 0.22 &(0.03) \\
    \hline
  \end{tabular} \\
  \begin{tabular}[h]{|lccc|}
    \hline
    process & $T_{sd}$[s] & $T_{4 \text{ digits}}$[s] & (\% fixed) \\
    \hline
    6g     & 6.19 & 6.81 &(1.37) \\
    2u4g   & 7.19 & 7.40 &(0.38) \\
    2u2d2g & 2.05 & 2.06 &(0.08) \\
    4u2g   & 4.08 & 4.15 &(0.21) \\
    2u2d2s & 0.38 & 0.38 &(0.00) \\
    2u4d   & 0.74 & 0.74 &(0.00) \\
    6u     & 2.16 & 2.17 &(0.02) \\
    \hline
  \end{tabular}
  \begin{tabular}[h]{|lccc|}
    \hline
    process & $T_{sd}$[s] & $T_{4 \text{ digits}}$[s] & (\% fixed) \\
    \hline
    7g       & 171.3 & 276.7 &(8.63) \\
    2u5g     & 195.1 & 241.2 &(3.25) \\
    2u2d3g   & 45.7  & 48.8  &(0.88) \\
    4u3g     & 92.5  & 101.5 &(1.29) \\
    2u2d2s1g & 7.9   & 8.1   &(0.23) \\
    2u4d1g   & 15.8  & 16.2  &(0.29) \\
    6u1g     & 47.1  & 48.6  &(0.41) \\
    \hline
  \end{tabular}
  \caption{Timing estimates in seconds for full colour and helicity summed virtual corrections.
  $T_{sd}$ is the time for evaluation in double precision, $T_{4 \text{ digits}}$ is the average
  time estimated to obtain a result correct to 4 digits using the phase space cuts of section
  \ref{sec:acc}. All times include the two evaluations necessary to obtain the accuracy estimate via
  the scaling test and were obtained on an Intel(R) Xeon(R) CPU E3-1240 @ 3.30GHz.}
  \label{tab:timechsum}    
  \end{center}
\end{table}

Table \ref{tab:timedsymgluons} shows the evaluation times using the de-symmetrized colour sums for
the pure gluonic channels which exploit the Bose symmetry of the final state. Though these channels
benefit the most from this treatment the complex channels with a single fermion pair would also
see a considerable speed up. Again we stress that these de-symmetrized sums contain the same full
colour information as the standard ones after the integration over the phase-space or equivalently
after explicitly summing the $\frac{1}{2}(n-2)!$ permutations of the final state gluons.

\begin{table}
  \centering
  \begin{tabular}{|c|c|c|c|c|}
    \hline
    & $gg\to2g$ & $gg\to3g$ & $gg\to4g$ & $gg\to5g$ \\
    \hline
    standard sum & 0.03 & 0.22 & 6.19 & 171.31 \\
    de-symmetrized & 0.03 & 0.07 & 0.57 & 3.07 \\
    \hline
  \end{tabular}
  \caption{Timing estimates in seconds for the de-symmetrized colour and helicity summed gluonic channels.
   All times include the two evaluations necessary to obtain the accuracy
  estimate via the scaling test and were obtained on an Intel(R) Xeon(R) CPU E3-1240 @ 3.30GHz.}
  \label{tab:timedsymgluons}
\end{table}

\section{Conclusions}

In this paper we have presented the C++ library \NJet for the numerical evaluation of one-loop
amplitudes for multi-jet production at hadron colliders. Using generalized unitarity 
together with off-shell recursion relations we were able to construct multi-fermion primitive
amplitudes at a computational cost growing polynomially with time. Accuracy estimates obtained by
applying a momentum re-scaling test are used to ascertain when higher precision numerics are
required to achieve a preset number of correct digits. A simple
cache system allows the efficient computation of helicity and permutation sums.

In order to use the multi-fermion primitive amplitudes for NLO QCD computations the colour
space information must be included. By matching primitive and partial amplitudes via Feynman
diagrams we obtained full colour formulae for all channels contributing to $\leq 5$ jet production.
We employed Furry type symmetries and linear algebra techniques to minimize the number of primitive amplitudes
appearing in the colour sums. Furthermore we made use of Bose symmetries in the final state to simplify
the most complex all-gluon processes. Since the on-shell algorithm is particularly well suited to
the computation of multi-gluon states which contain degenerate degrees of freedom we find a
favourable performance in comparison to existing public one-loop amplitude generators.

The interface to the library is either directly to the primitive amplitudes using the {\tt NParton}
class or via a Binoth Les Houches Accord \cite{Binoth:2010xt} to the colour and helicity summed
amplitudes. The latter allows the library to be linked  with
existing parton level Monte Carlo programs which can combine the
virtual and real radiation contributions 
to obtain full NLO
predictions.
The package presented here has been recently used to calculate three-
and four jet cross sections at next-to-leading order in QCD
\cite{Badger:2012zz}. This application nicely illustrates that the tool
presented can be used for state of the art calculations. We also
provide in Table~\ref{tab:7parton} in the appendix benchmark points for all  one-loop corrections required for
the evaluation of the five jet cross section at next-to-leading
order accuracy. 

Future developments to include heavy quark corrections would be
interesting. 

The library \NJet is publicly available from the project home page at 
 \url{https://bitbucket.org/njet/njet}.

\acknowledgments{We are grateful to Nigel Glover, Steven Wells, Rikkert Frederix and Marco Zaro for testing
preliminary versions of the code.This work is supported by the
Helmholtz Gemeinschaft under contract HA-101 (Alliance Physics at the
Terascale) and by the European Commission through contract
PITN-GA-2010-264564 (LHCPhenoNet).}

\begin{appendix}

\section{Numerical cross checks \label{app:numchecks}}

With \verb1--enable-demo1 used to configure the package two example programs are compiled into
their respective source directories:
\begin{verbatim}
  src-tools/NJet-demo
  src-nparton/NParton-demo
\end{verbatim}
\verb1NParton-demo1 is a short program showing how to use the \verb1NParton1 library by evaluating
some one-loop multi-parton primitive amplitudes. The code found in \verb1src-nparton/NParton-demo.cpp1
has some comments to explain the functions calls.

A number of hard coded cross checks for the colour summed amplitudes can be found in the
program \verb1NJet-demo1. The individual checks can be accessed with command line options, simply
type \verb1NJet-demo --help1 to see the available options.

For convenience we list the program output for the comparison with {\sc HELAC-1LOOP}.

\verb2NJet-demo --Helac1l2 gives a numerical cross check with {\sc HELAC-1LOOP}~\cite{Bevilacqua:2011xh}
for the available channels of dijet, three-jet and four-jet
production. We find again good numerical agreement. The exact phase-space points for these checks are
displayed to double precision when running the program. The results of these checks are as follows:
\begin{verbatim}

==== 0 --> g  g  g  g ====
          NJet:                     Helac1loop:               rel.diff.
born:     +1.4064482826874553e+02,  +1.4064482826874587e+02,  +2.4e-15
1/eps^2:  -2.7963100376599282e+01,  -2.7963100376599272e+01,  +3.8e-16
1/eps:    -2.4893580169045921e+00,  -2.4893580169051024e+00,  +2.0e-13
finite:   +2.2538771232905535e+01,  +2.2538771232905539e+01,  +1.6e-16

==== 0 --> u~ u  g  g ====
          NJet:                     Helac1loop:               rel.diff.
born:     +3.8973755683505198e+00,  +3.8973755683505242e+00,  +1.1e-15
1/eps^2:  -5.5963473237272909e-01,  -5.5963473237273020e-01,  +2.0e-15
1/eps:    -2.2701782897620162e-01,  -2.2701782897620176e-01,  +6.1e-16
finite:   +1.5812149573186310e-01,  +1.5812149573186496e-01,  +1.2e-14

==== 0 --> u~ u  d~ d ====
          NJet:                     Helac1loop:               rel.diff.
born:     +5.5757038821628091e-01,  +5.5757038821628113e-01,  +4.0e-16
1/eps^2:  -4.9269566071365262e-02,  -4.9269566071365241e-02,  +4.2e-16
1/eps:    -4.9621618996857207e-02,  -4.9621618996857221e-02,  +2.8e-16
finite:   -2.8732424653763196e-02,  -2.8732424653763248e-02,  +1.8e-15

==== 0 --> u~ u  u~ u ====
          NJet:                     Helac1loop:               rel.diff.
born:     +5.4442906439318520e+01,  +5.4442906439318527e+01,  +1.3e-16
1/eps^2:  -4.8108336321631899e+00,  -4.8108336321631695e+00,  +4.2e-15
1/eps:    +3.8870984690869954e+00,  +3.8870984690871220e+00,  +3.3e-14
finite:   +2.3627403557985861e+01,  +2.3627403557985797e+01,  +2.7e-15

--------- 2 --> 3 --------------

==== 0 --> g  g  g  g  g ====
          NJet:                     Helac1loop:               rel.diff.
born:     +5.0536873228500680e-02,  +5.0536873228500589e-02,  +1.8e-15
1/eps^2:  -1.2559719367246716e-02,  -1.2559719367246622e-02,  +7.5e-15
1/eps:    -6.5137043664892195e-03,  -6.5137043664457188e-03,  +6.7e-12
finite:   +1.9628928574388416e-02,  +1.9628928574411668e-02,  +1.2e-12

==== 0 --> u~ u  g  g  g ====
          NJet:                     Helac1loop:               rel.diff.
born:     +5.0328282104265365e-04,  +5.0328282104265484e-04,  +2.4e-15
1/eps^2:  -9.7283503961992726e-05,  -9.7283503961991818e-05,  +9.3e-15
1/eps:    -6.1775045952609201e-05,  -6.1775045952603048e-05,  +1.0e-13
finite:   +1.1409871173998674e-04,  +1.1409871173998645e-04,  +2.5e-15

==== 0 --> u~ u  d~ d  g ====
          NJet:                     Helac1loop:               rel.diff.
born:     +3.6329475567709317e-04,  +3.6329475567709729e-04,  +1.1e-14
1/eps^2:  -5.0160076608990277e-05,  -5.0160076608990379e-05,  +2.0e-15
1/eps:    -4.1238301409751345e-05,  -4.1238301409752870e-05,  +3.7e-14
finite:   +1.6929367019718101e-05,  +1.6929367019719798e-05,  +1.0e-13

==== 0 --> u~ u  u~ u  g ====
          NJet:                     Helac1loop:               rel.diff.
born:     +1.6607841328194570e-03,  +1.6607841328194737e-03,  +1.0e-14
1/eps^2:  -2.2930432666983780e-04,  -2.2930432666983907e-04,  +5.6e-15
1/eps:    -1.6167755068577280e-04,  -1.6167755068572718e-04,  +2.8e-13
finite:   +7.0356660045273211e-04,  +7.0356660045274610e-04,  +2.0e-14

--------- 2 --> 4 --------------

==== 0 --> u~ u  g  g  g  g ====
          NJet:                     Helac1loop:               rel.diff.
born:     +2.4749786907729620e-05,  +2.4749786907729492e-05,  +5.2e-15
1/eps^2:  -6.0142737839968278e-06,  -6.0142737839968761e-06,  +8.0e-15
1/eps:    -1.4238548051572628e-05,  -1.4238548051576113e-05,  +2.4e-13
finite:   -2.1818575805881747e-06,  -2.1818575805944554e-06,  +2.9e-12

==== 0 --> u~ u  d~ d  g  g ====
          NJet:                     Helac1loop:               rel.diff.
born:     +3.8990501849846650e-06,  +3.8990501849846963e-06,  +8.0e-15
1/eps^2:  -7.3214448592467342e-07,  -7.3214448592467173e-07,  +2.3e-15
1/eps:    -1.5000972850719066e-06,  -1.5000972850719574e-06,  +3.4e-14
finite:   +1.1570207960122947e-07,  +1.1570207960076906e-07,  +4.0e-12

==== 0 --> u~ u  u~ u  g  g ====
          NJet:                     Helac1loop:               rel.diff.
born:     +1.2029972817541046e-05,  +1.2029972817541090e-05,  +3.7e-15
1/eps^2:  -2.2589291869350247e-06,  -2.2589291869350408e-06,  +7.1e-15
1/eps:    -4.9249280446345875e-06,  -4.9249280446538397e-06,  +3.9e-12
finite:   +3.6301807851586739e-06,  +3.6301807851561252e-06,  +7.0e-13

==== 0 --> u~ u  d~ d  s~ s ====
          NJet:                     Helac1loop:               rel.diff.
born:     +4.6602437519749336e-08,  +4.6602437519749931e-08,  +1.3e-14
1/eps^2:  -6.1770188741857187e-09,  -6.1770188741856584e-09,  +9.8e-15
1/eps:    -1.2079244035518674e-08,  -1.2079244035518879e-08,  +1.7e-14
finite:   +3.0749739659342438e-09,  +3.0749739659321969e-09,  +6.7e-13

==== 0 --> u~ u  d~ d  d~ d ====
          NJet:                     Helac1loop:               rel.diff.
born:     +1.7752256666856891e-08,  +1.7752256666857116e-08,  +1.3e-14
1/eps^2:  -2.3530104931548807e-09,  -2.3530104931548365e-09,  +1.9e-14
1/eps:    -4.8669803368255748e-09,  -4.8669803368265401e-09,  +2.0e-13
finite:   +1.2351513905489450e-09,  +1.2351513905470286e-09,  +1.6e-12

==== 0 --> u~ u  u~ u  u~ u ====
          NJet:                     Helac1loop:               rel.diff.
born:     +1.9800262627817160e-06,  +1.9800262627817295e-06,  +6.8e-15
1/eps^2:  -2.6244677848455005e-07,  -2.6244677848455651e-07,  +2.5e-14
1/eps:    -3.5478617113255096e-07,  -3.5478617113278930e-07,  +6.7e-13
finite:   +1.2984349868781561e-06,  +1.2984349868773965e-06,  +5.8e-13
\end{verbatim}

\section{Numerical evaluation for \texorpdfstring{$7$}{7}-parton virtual corrections \label{app:7parton}}

We evaluate all indepenent channels contributing to the virtual corrections to $pp\to 5$ jets
at the following phase space point:
\begin{gather}
\label{eq:7g-pspoint}
\begin{aligned}
p_1 = \bigl\{&{-}5.0000000000000000 \times 10^{2},\; 0.0000000000000000 \times 10^{0},\\
             &0.0000000000000000 \times 10^{0},\; -5.0000000000000000 \times 10^{2}\bigr\},\\
p_2 = \bigl\{&{-}5.0000000000000000 \times 10^{2},\; 0.0000000000000000 \times 10^{0},\\
             &0.0000000000000000 \times 10^{0},\; 5.0000000000000000 \times 10^{2}\bigr\},\\
p_3 = \bigl\{&8.6354068143781365 \times 10^{1},\; -1.5213389320261800 \times 10^{1},\\
             &3.7633551294916273 \times 10^{1},\; -7.6218722682185415 \times 10^{1}\bigr\},\\
p_4 = \bigl\{&2.8011818180937634 \times 10^{2},\; -8.3126111650582232 \times 10^{1},\\
             &{-}2.6320385675865049 \times 10^{2},\; 4.7749085116026578 \times 10^{1}\bigr\},\\
p_5 = \bigl\{&1.2752252956966052 \times 10^{2},\; -9.0449041295993482 \times 10^{1},\\
             &{-}8.3178307703078929 \times 10^{1},\; 3.4093043339258053 \times 10^{1}\bigr\},\\
p_6 = \bigl\{&4.1413006837454346 \times 10^{2},\; 2.3214556494593859 \times 10^{2},\\
             &3.3275443678081871 \times 10^{2},\; -8.2985751852442578 \times 10^{1}\bigr\},\\
p_7 = \bigl\{&9.1875152102638314 \times 10^{1},\; -4.3357022679101075 \times 10^{1},\\
             &{-}2.4005823614005564 \times 10^{1},\; 7.7362346079343363 \times 10^{1}\bigr\}.
\end{aligned}
\end{gather}
The result has been obtained with the number of light quark flavour $n_f=5$, a renormalization
scale of $\mu_R=91.188$ and $\alpha_s=0.118$. Table \ref{tab:7parton} shows the numerical values of the amplitudes.
\begin{table}[h]
\begin{tabular}{|c|c|c|}
  \hline
  Channel & Born & $2\operatorname{Re} (\text{Loop}\cdot \text{Born})$ \\
  \hline
  \multirow{3}{*}{$ 0 \to 7\,g$}
                                 &                                & $-2.6321266028904 \times10^{-7} \cdot \epsilon^{-2}$ \\
                                 & $6.6739867626785 \times10^{-7}$ & $ 1.2781291277818 \times10^{-7} \cdot \epsilon^{-1}$ \\
                                 &                                & $-4.0556181173518 \times10^{-7} \cdot \epsilon^{0}$  \\
  \hline
  \multirow{3}{*}{$ 0 \to \bar u u + 5\,g$}
                                 &                                & $-2.7277340128627 \times10^{-9} \cdot \epsilon^{-2}$ \\
                                 & $8.2213902805466 \times10^{-9}$ & $-4.6877933000518 \times10^{-11} \cdot \epsilon^{-1}$ \\
                                 &                                & $-4.8319141582166 \times10^{-9} \cdot \epsilon^{0}$  \\
  \hline
  \multirow{3}{*}{$ 0 \to \bar u u \bar d d + 3\,g$}
                                 &                                & $-6.9151881689636 \times 10^{-10} \cdot \epsilon^{-2}$ \\
                                 & $2.5689441485779 \times10^{-9}$ & $ 7.9158077535594 \times 10^{-11} \cdot \epsilon^{-1}$ \\
                                 &                                & $-1.9006336322439 \times 10^{-9} \cdot \epsilon^{0}$  \\
  \hline
  \multirow{3}{*}{$ 0 \to \bar u u \bar u u + 3\,g$}
                                 &                                & $-1.2283346215940 \times 10^{-9} \cdot \epsilon^{-2}$ \\
                                 & $4.5631774024631 \times10^{-9}$ & $ 4.9063178156336 \times 10^{-11} \cdot \epsilon^{-1}$ \\
                                 &                                & $-1.9994502250899 \times 10^{-9} \cdot \epsilon^{0}$  \\
  \hline
  \multirow{3}{*}{$ 0 \to \bar u u \bar d d \bar s s + g$}
                                 &                                & $-5.9292119796938 \times 10^{-12} \cdot \epsilon^{-2}$ \\
                                 & $2.8701338670235 \times10^{-11}$& $ 5.8421068466114 \times 10^{-13} \cdot \epsilon^{-1}$ \\
                                 &                                & $-1.5312833372355 \times 10^{-11} \cdot \epsilon^{0}$  \\
  \hline
  \multirow{3}{*}{$ 0 \to \bar u u \bar d d \bar d d + g$}
                                 &                                & $-2.6400462631767 \times 10^{-12} \cdot \epsilon^{-2}$ \\
                                 & $1.2779583891422 \times10^{-11}$& $-4.5891042883903 \times 10^{-13} \cdot \epsilon^{-1}$ \\
                                 &                                & $-6.6633865124667 \times 10^{-12} \cdot \epsilon^{0}$  \\
  \hline
  \multirow{3}{*}{$ 0 \to \bar u u \bar u u \bar u u + g$}
                                 &                                & $-3.9440369566465 \times 10^{-10} \cdot \epsilon^{-2}$ \\
                                 & $1.9091768148671 \times10^{-9}$ & $ 1.9659510267642 \times 10^{-10} \cdot \epsilon^{-1}$ \\
                                 &                                & $-3.6659249528662 \times 10^{-10} \cdot \epsilon^{0}$  \\
  \hline
\end{tabular}
\caption{The numerical evaluation of 7 parton virtual corrections at the phase space point
 given in eq.~\eqref{eq:7g-pspoint}. Using $\alpha_s = 0.118$, $\mu_R = 91.188$
 and including final state symmetry factors.
\label{tab:7parton}
  }
\end{table}
\normalsize

\end{appendix}

\providecommand{\href}[2]{#2}\begingroup\raggedright\endgroup

\end{document}